\documentclass[lettersize,journal]{IEEEtran}
\usepackage{amsmath,amsfonts}
\usepackage{array}
\usepackage{tikz}
\usepackage{amsmath,mathtools}
\usepackage{amssymb}
\usepackage{booktabs}
\usepackage{graphicx}
\usepackage{amsmath}
\usepackage{algorithm}
\usepackage{algpseudocode}
\usepackage{boxedminipage}
\usepackage{fancyhdr}
\usepackage{multirow}
\usepackage{mathptmx}
\usepackage{array}
\usepackage{textcomp}
\usetikzlibrary{positioning}
\usetikzlibrary{arrows.meta}
\usetikzlibrary{calc}
\usetikzlibrary{positioning, fit, calc, arrows.meta, backgrounds}
\usetikzlibrary{shapes.geometric}
\usetikzlibrary{fit}
\usepackage{stfloats}
\usepackage{csquotes} 
\usepackage{url}
\usepackage{verbatim}
\usepackage{cite}
\usepackage{eqnarray}
\usepackage[caption=false,font=normalsize,labelfont=sf,textfont=sf]{subfig}
\usepackage{textcomp}
\usepackage{color}
\usepackage{stfloats}
\usepackage{url}
\usepackage{verbatim}
\usepackage{graphicx}
\def\BibTeX{{\rm B\kern-.05em{\sc i\kern-.025em b}\kern-.08em
    T\kern-.1667em\lower.7ex\hbox{E}\kern-.125emX}}
\usepackage{balance}
\usepackage{pifont}

\newcommand{\xmark}{\ding{55}}

\begin{document}
\title{Intelligent 6G Edge Connectivity: A Knowledge Driven Optimization Framework for Small Cell Selection}

\author{
Tuğçe Bilen, \textit{Member, IEEE}, Ian F. Akyildiz, \textit{Life Fellow, IEEE}
\thanks{
Tuğçe Bilen is with the Department of Artificial Intelligence and Data Engineering, 
Faculty of Computer and Informatics, 
Istanbul Technical University, Istanbul, Turkey 
(e-mail: bilent@itu.edu.tr).}
\thanks{
Ian F. Akyildiz is with Truva Inc., Alpharetta, GA 30022, USA 
(e-mail: ian@truvainc.com).}
}

\markboth{submitted to Transactions on Mobile Computings,~Vol.~xx, No.~xx, February~2026}{}
%{How to Use the IEEEtran \LaTeX \ Templates}

\maketitle

\begin{abstract}
Sixth-generation (6G) wireless networks are expected to support immersive and mission-critical applications requiring ultra-reliable communication, low latency, and high data rates. Dense small-cell deployments enable these capabilities; however, the large number of candidate cells significantly complicates user–cell association. Conventional signal-strength-based or heuristic schemes often cause load imbalance, increased latency, packet loss, and inefficient resource utilization. To address this problem, this paper proposes a Knowledge-Defined Networking (KDN) framework for intelligent user association in dense 6G small-cell networks. The architecture integrates the knowledge, control, and data planes to support adaptive decision-making. Small-cell conditions are represented using queue-aware indicators capturing traffic load and waiting-time dynamics. Based on these indicators, a joint latency–packet-loss objective is formulated and solved via Lagrangian relaxation to obtain globally guided association decisions. The resulting decisions are then used to train a lightweight Learning Vector Quantization (LVQ) model, enabling fast inference at the network edge. Extensive NS-3 simulations under varying mobility, traffic load, packet size, and network density show that the proposed framework consistently outperforms baseline methods. The results indicate latency reductions of 30–45\% in high-mobility and heavy-traffic scenarios and packet loss reductions exceeding 35\% under congestion. These findings demonstrate that combining optimization-driven knowledge with lightweight learning enables scalable and QoS-aware user association in dense 6G networks.
\end{abstract}


\section{Introduction}

Sixth-generation (6G) wireless networks are expected to support an unprecedented range of applications that demand extremely stringent performance guarantees. Emerging services such as immersive extended reality, connected autonomous systems, large-scale Internet of Things (IoT), and real-time digital twin platforms require not only multi-gigabit data rates but also consistently low latency, high reliability, and improved energy efficiency \cite{9749230}. Meeting these requirements necessitates a fundamental evolution in network architecture and resource management strategies \cite{BILEN2026111941}.

\begin{figure}[h]
\centering
\includegraphics[width=0.4\textwidth]{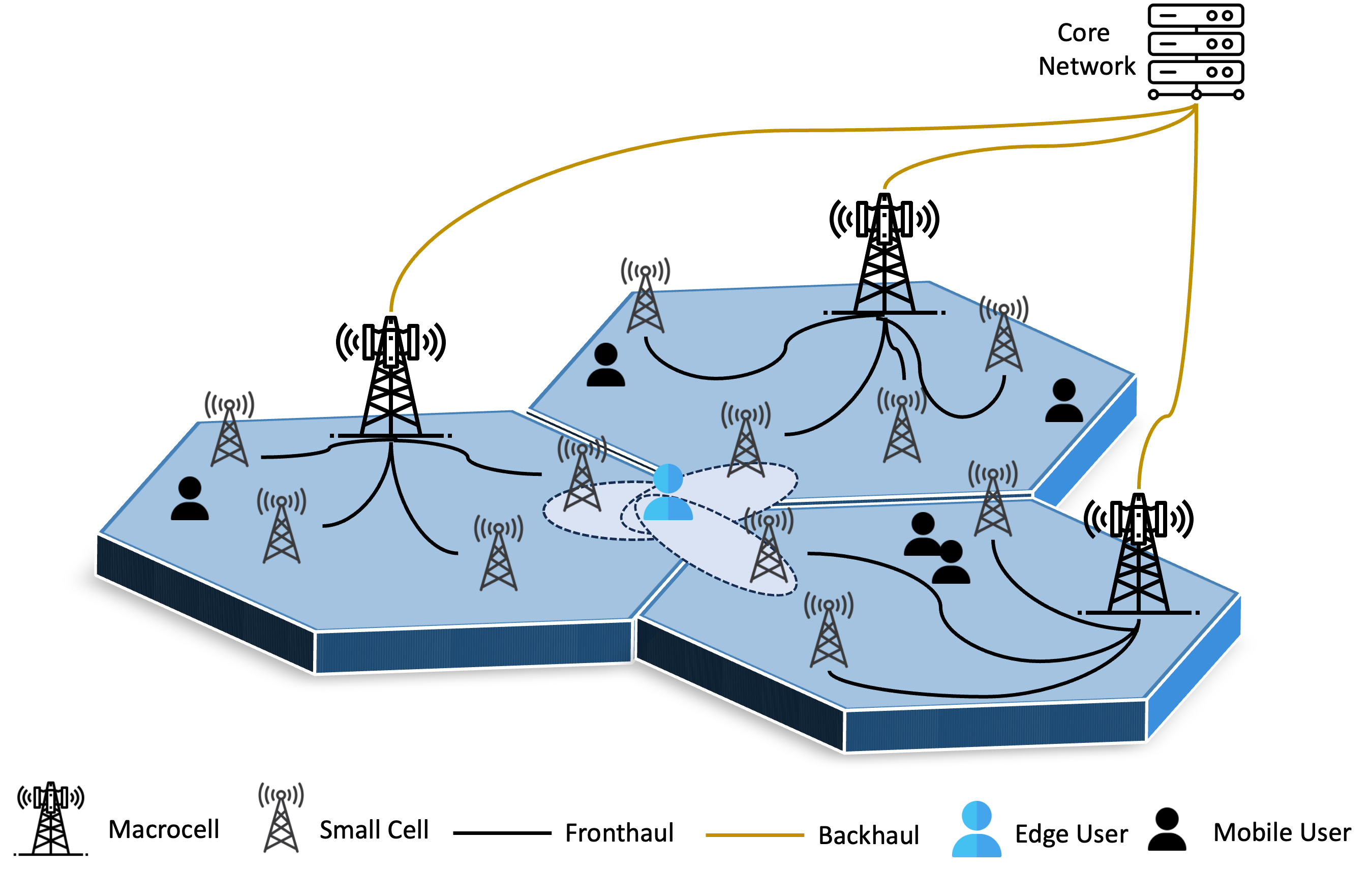}
\caption{Illustration of dense 6G small-cell architecture and the multi-association challenge encountered by edge users.}
\label{fig1}
\end{figure}

One of the key enablers envisioned for 6G is the large-scale deployment of heterogeneous small cells operating across diverse spectrum bands, including millimeter-wave and terahertz frequencies \cite{8782879}. Network densification significantly increases spatial reuse and enhances aggregate network capacity. However, it also introduces substantial challenges in mobility management, interference coordination, and user association \cite{7084385}. In dense environments, users are frequently surrounded by multiple candidate cells, each offering different radio conditions and traffic states. Determining which cell should serve a user, therefore, becomes a non-trivial decision that directly impacts end-to-end service quality.

Traditional cell selection mechanisms are primarily based on radio-layer indicators such as received signal strength or signal-to-interference-plus-noise ratio. While such metrics are effective in sparse deployments, they become insufficient in ultra-dense networks. In practice, cells with similar signal quality can experience drastically different traffic loads, queueing delays, and scheduling conditions. As a result, associating a user with the strongest signal does not necessarily lead to the best service experience. A user may be connected to a congested cell even though neighboring cells with slightly weaker signal conditions could offer significantly lower delay and higher reliability. Consequently, user association in dense 6G networks must evolve from a purely signal-driven mechanism to a network-aware decision process that incorporates service dynamics alongside radio measurements. This observation motivates a shift from purely signal-driven association to knowledge-driven decision mechanisms that jointly consider radio conditions and service-level dynamics.

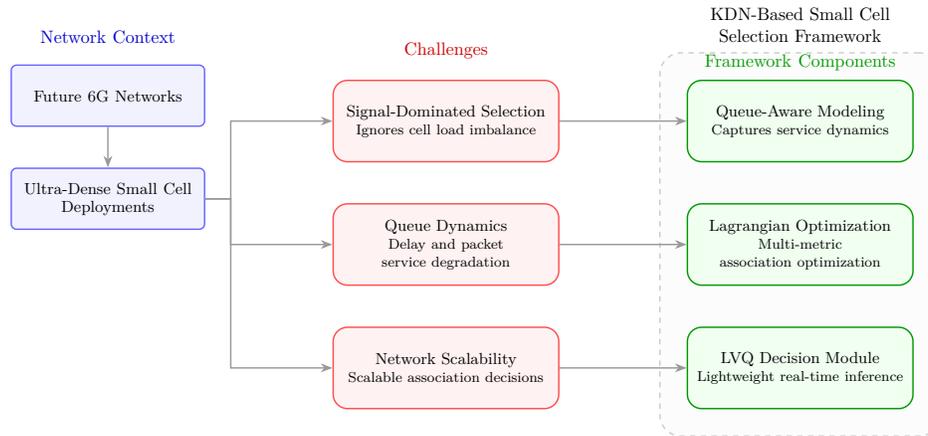
\begin{figure*}[h]
\centering
		\resizebox{0.7\linewidth}{!}{%
\begin{tikzpicture}[
    node distance=0.8cm and 2.0cm,
    font=\sffamily\small,
    context/.style={rectangle, draw=blue!60, fill=blue!5, thick, rounded corners=3pt, minimum width=3.8cm, align=center, minimum height=1.2cm},
    prob/.style={rectangle, draw=red!70, fill=red!5, thick, rounded corners=8pt, text width=4.2cm, align=center, minimum height=1.6cm},
    sol/.style={rectangle, draw=green!60!black, fill=green!6, thick, rounded corners=8pt, text width=4.2cm, align=center, minimum height=1.6cm},
    group/.style={draw=gray!40, dashed, thick, rounded corners=12pt, fill=gray!2},
    arrow/.style={-Stealth, thick, gray!80}
]

% Context
\node[context] (sixg) {{Future 6G Networks}};
\node[context, below=of sixg] (dense) {Ultra-Dense Small Cell\\Deployments};

% Challenges
\node[prob, right=2.5cm of sixg, yshift=-0.5cm] (c1)
{\textbf{Signal-Dominated Selection}\\
\footnotesize Ignores cell load imbalance};

\node[prob, below=of c1] (c2)
{\textbf{Queue Dynamics}\\
\footnotesize Delay and packet service degradation};

\node[prob, below=of c2] (c3)
{\textbf{Network Scalability}\\
\footnotesize Scalable association decisions};

% Framework components
\node[sol, right=2.5cm of c1] (s1)
{\textbf{Queue-Aware Modeling}\\
\footnotesize Captures service dynamics};

\node[sol, right=2.5cm of c2] (s2)
{\textbf{Lagrangian Optimization}\\
\footnotesize Multi-metric association optimization};

\node[sol, right=2.5cm of c3] (s3)
{\textbf{LVQ Decision Module}\\
\footnotesize Lightweight real-time inference};

\begin{scope}[on background layer]
\node[group, fit={(s1) (s3)}, inner sep=15pt] (fwbox) {};
\end{scope}

\node[anchor=south, yshift=2pt, font=\bfseries\sffamily, align=center, text width=6cm]
at (fwbox.north) {KDN-Based Small Cell\\Selection Framework};

\node[above=3mm of sixg, blue!80!black, font=\scshape] {Network Context};
\node[above=3mm of c1, red!80!black, font=\scshape] {Challenges};
\node[above=0.6mm of s1, green!60!black, font=\scshape] {Framework Components};

% Arrows
\draw[arrow] (sixg) -- (dense);
\draw[arrow] (dense.east) -- ++(0.5,0) |- (c1.west);
\draw[arrow] (dense.east) -- ++(0.5,0) |- (c2.west);
\draw[arrow] (dense.east) -- ++(0.5,0) |- (c3.west);

\draw[arrow] (c1.east) -- (s1.west);
\draw[arrow] (c2.east) -- (s2.west);
\draw[arrow] (c3.east) -- (s3.west);

\end{tikzpicture}
		}%
\caption{Conceptual illustration of the challenges in dense 6G deployments and the main components of the proposed KDN-enabled small-cell selection framework.}
\label{fig:framework-overview}
\end{figure*}

The problem becomes even more critical for edge users located within the overlapping coverage regions of multiple small cells, as illustrated in Fig. \ref{fig1}. These users are particularly sensitive to association decisions because their achievable performance strongly depends on the selected serving cell. Although frequent handovers are undesirable, remaining attached to a heavily loaded cell for extended periods may lead to persistent service degradation. This observation highlights the need for intelligent association strategies that jointly consider channel quality, traffic conditions, and service requirements.

At the same time, the scale and operational complexity of future 6G infrastructures make manual configuration and static rule-based management increasingly impractical. Knowledge-Defined Networking (KDN) has recently emerged as a promising paradigm for enabling intelligent and adaptive network operation \cite{11357882}. By integrating data analytics, machine learning, and programmable control into a closed-loop architecture, KDN allows networks to continuously observe their state, extract operational knowledge, and apply informed control decisions. Such capabilities are particularly valuable in dense small-cell environments, where network conditions evolve rapidly.

Motivated by these challenges, this paper develops a KDN-enabled framework for intelligent small-cell selection in dense 6G networks. As summarized in Fig.~\ref{fig:framework-overview}, the proposed framework consists of three tightly coupled components. First, a queue-aware modeling layer captures the service dynamics of small cells by incorporating traffic load and delay indicators. Second, a multi-metric association problem is formulated and solved using a tractable Lagrangian relaxation-based optimization framework to balance latency, packet service degradation, and energy consumption under cell capacity constraints. Finally, a lightweight LVQ-based learning module, chosen for its low computational overhead and stable behavior in edge environments, approximates the resulting policies and enables real-time decision making. The main contributions of this work can be summarized as follows:

\begin{itemize}

\item {Queue-aware modeling of small-cell service dynamics that reflect actual service conditions rather than solely radio signal strength:}
We develop a modeling framework capturing the interaction between traffic load, service delay, and packet handling characteristics, enabling service-aware association decisions.

\item {Lagrangian-based multi-metric association optimization:}
We formulate the small-cell selection problem as a constrained optimization task that jointly considers latency effects, packet service degradation, and energy consumption, and solve it using a Lagrangian relaxation-based approach.

\item {Knowledge-Defined Networking control architecture:}
We design a KDN-driven closed-loop framework that integrates data collection, knowledge extraction, and automated dissemination of association decisions.

\item {Learning-assisted real-time inference:}
We incorporate a lightweight LVQ-based learning module to approximate optimization outcomes, enabling scalable and efficient decision making in dynamic network environments.

\end{itemize}

The remainder of this paper is organized as follows. Section~2 reviews related studies in intelligent user association and learning-assisted network management. Section~3 presents the system model and the proposed KDN architecture. Section~4 details the optimization formulation and the learning-assisted decision process. Section~5 evaluates the performance of the proposed framework through extensive simulations. Finally, Section~6 concludes the paper and Section~7 outlines potential directions for future research.

\begin{table*}[t]
\centering
\caption{Comparison of Existing Studies and the Proposed Framework}
\label{tab:lit_comparison}
        \renewcommand{\arraystretch}{1}

\begin{tabular}{lcccccc}
\toprule
\textbf{Study Category} & 
\textbf{Radio Conditions} & 
\textbf{Load Awareness} & 
\textbf{Delay Modeling} & 
\textbf{Packet Loss} & 
\textbf{Learning / Intelligence} & 
\textbf{KDN Integration} \\
\midrule

Radio-centric approaches 
& \checkmark & \xmark & \xmark & \xmark & \xmark & \xmark \\

Optimization-based methods 
& \checkmark & \checkmark & Partial & \xmark & Limited & \xmark \\

Learning-based association 
& \checkmark & \checkmark & Limited & \xmark & \checkmark & \xmark \\

KDN-based network control 
& Partial & Partial & \xmark & \xmark & \checkmark & \checkmark \\

\midrule

\textbf{Proposed Framework}
& \checkmark & \checkmark & \checkmark & \checkmark & \checkmark & \checkmark \\

\bottomrule
\end{tabular}
\end{table*}

\section{Related Works}

This section critically reviews the literature most relevant to small cell selection and knowledge-driven network management. Existing studies can broadly be categorized into three groups: radio-centric association strategies, optimization-based approaches, and knowledge-driven intelligent control frameworks. While each line of work contributes valuable insights, none simultaneously considers radio conditions, traffic load dynamics, analytical delay modeling, packet loss awareness, and knowledge-driven adaptation within a unified architecture. This gap motivates the framework proposed in this study.

\subsection{Radio-Centric Small Cell Selection Approaches}

Early cell selection mechanisms primarily rely on radio indicators such as received signal strength, path loss, or coverage conditions. In \cite{7136539}, positioning information is combined with signal strength to improve association in ultra-dense deployments; however, network congestion and service-level performance are not considered. Similarly, \cite{7293389} proposes a handover mechanism based on path-loss variation trends, which enhances mobility robustness but remains purely radio-driven.

A network information service–assisted approach is introduced in \cite{7056313}, using coverage and distance information to support association decisions. Although this improves coverage consistency, it lacks visibility into queue dynamics and delay conditions. Mobility-aware reselection using vehicle trajectory information is investigated in \cite{9411865}, yet the objective focuses on connectivity rather than service performance. In \cite{8904157}, an uplink measurement-based user-centric scheme reduces measurement overhead but does not incorporate network load or delay-sensitive requirements. Likewise, the energy-efficient association framework in \cite{7447701} optimizes power consumption under backhaul constraints, while latency and packet loss remain outside the optimization objective.

Overall, these studies enhance radio efficiency, mobility robustness, or energy consumption, but largely overlook service-level quality indicators, limiting their suitability for latency-critical and reliability-sensitive 6G scenarios.

\subsection{Optimization-Based Approaches}

Optimization techniques have been widely applied to improve user association and load balancing in heterogeneous and ultra-dense networks. Several studies incorporate traffic load into association decisions and demonstrate gains in fairness and throughput \cite{8796404, 6497017}. However, load is often modeled as a long-term average metric, which does not fully capture instantaneous queue dynamics affecting latency and reliability. To support delay-sensitive services, cross-layer formulations have introduced latency-aware association and scheduling strategies \cite{10609965, birabwa2022service}. Although these approaches reduce delay, they frequently rely on computationally intensive or centralized optimization, limiting real-time applicability.

To improve scalability, recent work employs machine learning and policy approximation to learn association strategies under dynamic conditions \cite{zhao2018deep}. While these methods reduce online computational cost, they typically emphasize throughput or load balancing, and rarely integrate analytical delay models or packet loss estimation into the decision process.

Consequently, despite progress in optimization and learning-based strategies, the joint consideration of queue-aware delay modeling, packet loss behavior, and scalable real-time control remains insufficiently addressed in dense 6G small-cell networks.

\subsection{Knowledge-Defined Networking and Intelligent Control}

Knowledge-Defined Networking (KDN) extends the Software-Defined Networking paradigm by introducing a knowledge plane that exploits telemetry, analytics, and machine learning to enable adaptive network management. The concept was formalized in \cite{10.1145/3138808.3138810}, where learning-driven control loops were proposed for self-optimizing networks. Subsequent studies have explored intelligent decision mechanisms in KDN environments. For instance, \cite{10297317} applies graph neural networks and reinforcement learning for adaptive routing, while \cite{8406169} demonstrates how large-scale telemetry and analytics support self-driving networks. A cloud-native knowledge extraction and monitoring architecture is presented in \cite{10027809}, highlighting the potential of knowledge-centric network control.

Despite these advances, most KDN research focuses on core network routing, monitoring, or traffic engineering. Applications of KDN principles to radio access decisions, particularly user association and small-cell selection, remain limited. Furthermore, existing studies rarely combine analytical service models, such as delay and packet loss estimation, with knowledge-driven control mechanisms at the network edge.

To position our work within the literature, Table~\ref{tab:lit_comparison} summarizes key characteristics of prior approaches and contrasts them with the proposed framework. Radio-centric methods mainly emphasize signal quality and mobility robustness while overlooking traffic dynamics and service-level indicators. Optimization-based approaches improve load balancing and sometimes incorporate delay awareness, yet explicit packet loss modeling and knowledge-driven control are typically absent. Learning-based solutions enhance adaptability but often focus on throughput-oriented objectives rather than analytically grounded QoS metrics. KDN-oriented studies introduce intelligence into network management but largely focus on routing and monitoring rather than radio access control.

In contrast, the framework proposed in this paper integrates radio-condition awareness, queue-aware load modeling, analytical delay estimation, packet loss considerations, and knowledge-driven decision making within a unified KDN architecture. This design enables adaptive, service-aware small-cell selection that meets the stringent latency and reliability requirements anticipated in future dense 6G edge networks.

\section{Proposed System Model}
This section introduces the proposed KDN framework and formulates the small cell selection problem. First, the overall KDN-based architecture is described and the interaction among its main functional components is explained. Subsequently, the optimization objective and associated constraints are formally defined, forming the basis for the intelligent user association strategy proposed in this work.

\subsection{Proposed KDN-Based 6G Topology}
We consider a dense 6G deployment comprising multiple small cells operating within the coverage region of a macro cell, as illustrated in Fig.~\ref{fig2}. Such ultra-dense environments introduce highly dynamic traffic distributions, fluctuating radio conditions, and frequent user mobility, all of which make efficient user association a challenging problem. In these scenarios, traditional signal-strength-based association strategies are insufficient to ensure stable service quality.

To address these challenges, we adopt a KDN architecture composed of three logical planes: the Knowledge Plane, the Control Plane, and the Data Plane. These planes operate in a closed-loop manner, enabling continuous monitoring of network conditions, extraction of operational knowledge, and adaptive decision-making for small-cell selection \cite{BILEN2025103984}.

\begin{enumerate}
\item {Knowledge Plane:}
The knowledge plane constitutes the intelligence layer of the proposed framework. Its primary role is to collect network observations, transform raw measurements into structured knowledge, and support intelligent association decisions. The knowledge plane comprises three principal functional blocks: {Knowledge Collection and Generation}, {Knowledge Storage}, and a {Machine Learning (ML) Modules}.

\begin{itemize}
\item {Knowledge Collection and Generation Module:}  
This block aggregates raw monitoring data from the data plane, including traffic load indicators, radio measurements, queue statistics, and service performance metrics such as delay and packet loss. The collected measurements are processed and transformed into higher-level descriptors that characterize the service conditions of each small cell. These descriptors provide a more comprehensive representation of network state than traditional signal-based metrics alone.

\begin{figure}[h]
\centering
\includegraphics[width=0.75\columnwidth]{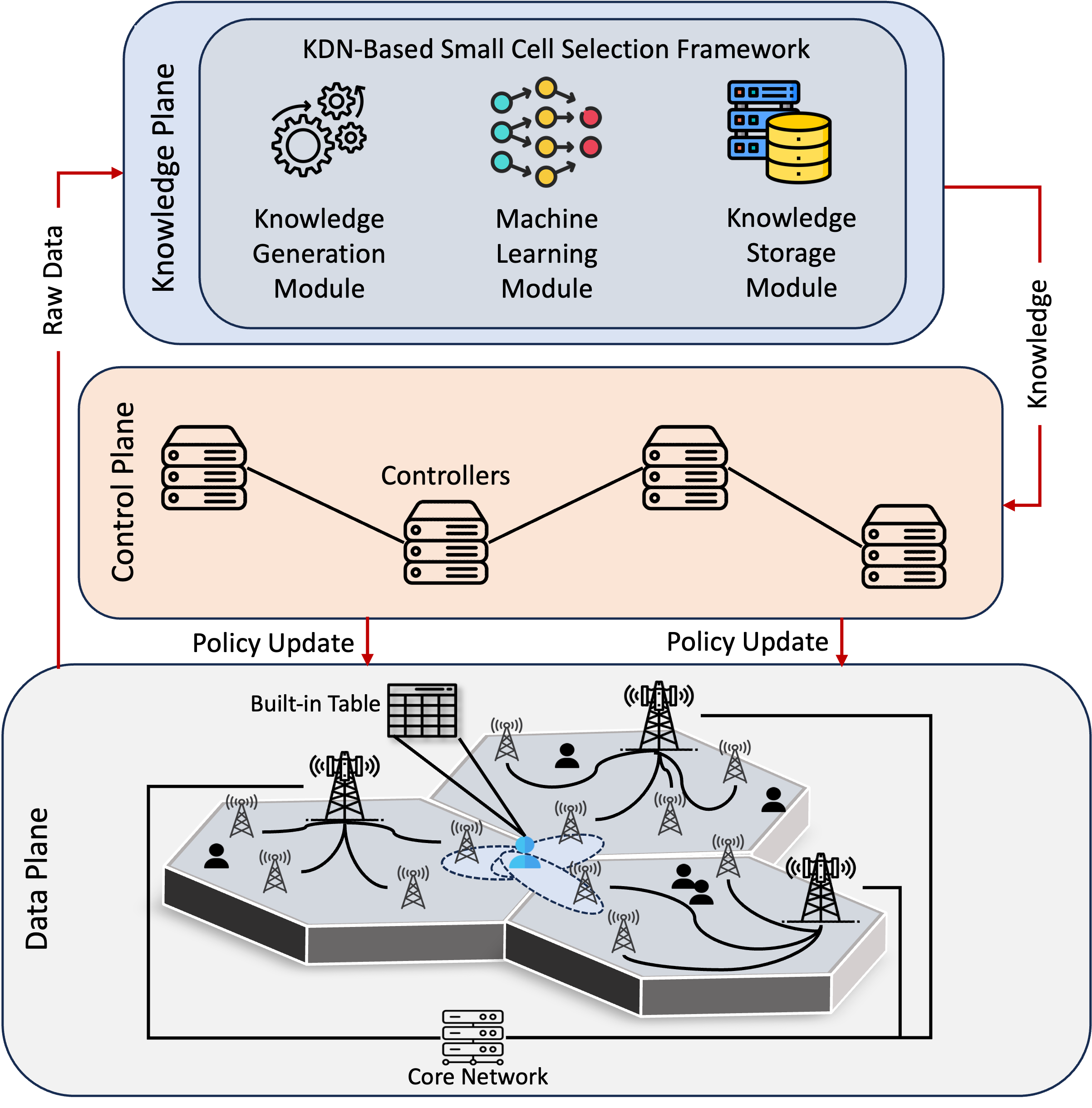}
\caption{Proposed KDN-Based 6G Topology}
\label{fig2}
\end{figure}

\item {Knowledge Storage Module:}  
The generated knowledge, together with historical observations, is persistently stored in a knowledge base. This repository enables the system to track temporal patterns in traffic demand, identify persistent congestion conditions, and support more informed and stable association decisions over time. By maintaining historical context, the knowledge base facilitates improved understanding of long-term network behavior.
\item {Machine Learning Module:}  
The ML module utilizes the stored knowledge to learn the complex relationship between observed network states and effective association decisions. Rather than replacing the optimization framework, the learning module serves as a decision-support mechanism that approximates suitable association policies with significantly lower computational overhead. This allows the system to react quickly to changing network conditions while still benefiting from the structure provided by the underlying optimization model.
\end{itemize}

\item {Control Plane:}
The control plane acts as the coordination layer between the knowledge and data planes. It receives high-level association decisions and policy updates generated by the knowledge plane and translates them into enforceable control actions. These actions include updating user-to-cell association tables, configuring network policies, and distributing control rules to the corresponding small cells. In this manner, the control plane ensures that knowledge-driven decisions are consistently applied across the entire network.

\item {Data Plane:}
The data plane represents the physical layer of the 6G system and includes user equipment (UEs), small cells, and macro cells. Small cells provide radio access and data services to users in accordance with policies received from the control plane. Each user maintains an association state indicating its currently serving small cell. In addition to providing connectivity, the data plane continuously reports operational measurements, such as channel quality indicators, traffic load levels, queue lengths, delay statistics, and packet loss rates, to the knowledge plane. These measurements close the control loop and allow the system to dynamically adapt association decisions as network conditions evolve.
\end{enumerate}

Through this structured interaction among the knowledge, control, and data planes, raw network measurements are transformed into actionable knowledge that supports intelligent and adaptive user association, as illustrated in Fig. \ref{fig2}. The resulting closed-loop architecture enables the network to continuously observe its operational state, reason about service performance, and dynamically adjust small-cell selection decisions in response to traffic fluctuations, mobility patterns, and varying radio conditions.

\vspace{-0.1in}
\subsection{Objective Function Formulation}

The objective of the proposed framework is to determine the most suitable serving small cell for each user at every decision epoch. In dense deployments, users are often located within the overlapping coverage of multiple cells. Although radio conditions may appear similar, internal traffic states can vary significantly due to differences in load, queue backlog, and scheduling behavior. As a result, association decisions based only on signal strength may lead to inefficient resource usage and degraded service quality.

To address this limitation, the framework evaluates user–cell associations using multiple service-related indicators that reflect each cell's operational condition. Rather than relying solely on physical-layer measurements, it incorporates traffic and service dynamics. Specifically, when user $j$ is associated with small cell $i$ at time slot $t$, three indicators are considered: the delay-related indicator $D_{ij}(t)$, the packet service degradation indicator $P_{ij}(t)$, and the energy-related indicator $E_{ij}(t)$. The delay indicator captures latency increases due to congestion and queue buildup, the packet service degradation indicator reflects performance deterioration such as drops or retransmissions under heavy load, and the energy indicator represents the relative communication cost between the user and the selected cell. These quantities are not direct measurements of delay, packet loss, or energy consumption; instead, they are normalized indicators derived from observable network statistics, enabling heterogeneous service effects to be integrated within a unified optimization framework.

Using these indicators, the objective is to minimize the aggregate service cost across users and decision intervals. Let $T$ denote the number of time slots, $M$ the number of users, and $N$ the number of candidate small cells. The objective function is given in Eq.~\ref{target}, where $\alpha$ is a non-negative parameter controlling the relative importance of the energy-related indicator.

\small
\begin{equation}\label{target}
J = \min_{X_{ij}(t)} \sum_{t=1}^{T} \sum_{j=1}^{M} \sum_{i=1}^{N}
\left( D_{ij}(t) X_{ij}(t) + P_{ij}(t) X_{ij}(t) + \alpha E_{ij}(t) X_{ij}(t) \right)
\end{equation}
\normalsize

The binary decision variable $X_{ij}(t)$ represents the association state between user $j$ and small cell $i$ at time slot $t$, as defined in Eq.~\ref{x}.

\small
\begin{equation} \label{x}
X_{ij}(t) =
\begin{cases}
1, & \text{if user } j \text{ is associated with small cell } i \text{ at time } t, \\
0, & \text{otherwise}
\end{cases}
\end{equation}
\normalsize

Two practical constraints govern the association process. First, each user can connect to only one serving cell at any time, expressed in Eq.~\ref{c1}. 

\small
\begin{equation} \label{c1}
\sum_{i=1}^{N} X_{ij}(t) = 1, \quad \forall j, \forall t
\end{equation}
\normalsize

Second, each cell has limited service capacity due to radio and processing constraints. Let $C_i(t)$ denote the maximum number of users supported by small cell $i$ at time $t$. The capacity constraint is given in Eq.~\ref{c2}.

\small
\begin{equation} \label{c2}
\sum_{j=1}^{M} X_{ij}(t) \leq C_i(t), \quad \forall i, \forall t
\end{equation}
\normalsize

Together, Eqs.~(\ref{target})–(\ref{c2}) define a constrained combinatorial optimization problem. The binary variables create a discrete solution space that grows rapidly with the number of users and cells, while the capacity constraint couples the decisions of multiple users sharing the same resources. Moreover, the delay and packet indicators depend on queue dynamics and traffic load, further complicating the objective structure.

Consequently, the problem is generally non-convex and belongs to the class of NP-hard assignment problems common in wireless resource allocation. Exhaustive search is therefore impractical in dense 6G scenarios. To enable scalable decision-making, the framework adopts a Lagrangian relaxation approach. By relaxing the capacity constraints and introducing Lagrange multipliers, the problem can be decomposed into tractable subproblems that can be solved efficiently. This provides a principled alternative to heuristic methods while preserving a close approximation to the original optimization structure. Under stable traffic conditions, the resulting dual formulation yields solutions with bounded deviation from the optimal primal solution. The next subsection presents the detailed definitions of the indicators $D_{ij}(t)$, $P_{ij}(t)$, and $E_{ij}(t)$ and explains how they are derived from queue and traffic dynamics within each small cell.

\begin{figure*}[h]
\centering
\includegraphics[width=0.69\textwidth]{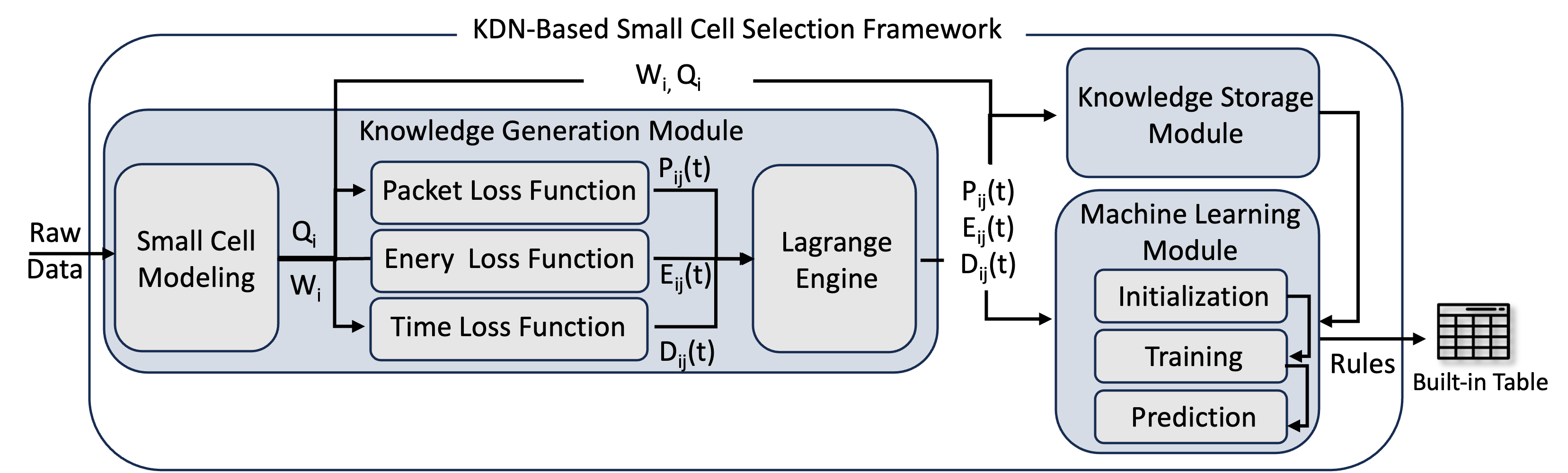}
\caption{KDN-Based Small Cell Selection Framework}
%\vspace{-2em}
\label{fig3}
\end{figure*}

\section{Proposed Small Cell Selection Approach}
This section introduces the proposed small-cell selection methodology within the KDN-based framework. The overall decision process operates within the Knowledge Plane and follows the architecture illustrated in Fig.~4. As previously described, this plane consists of three main components: the Knowledge Generation module, the Knowledge Storage module, and the Machine Learning module. The Knowledge Generation module derives association decisions from real-time network observations and analytical modeling. These decisions are then stored in the Knowledge Storage module, creating a historical repository of network states and corresponding optimal actions. Finally, the Machine Learning module utilizes this accumulated knowledge to approximate efficient decisions with minimal computational overhead during online operation. %The following subsections first explain how knowledge is generated through system modeling and optimization, and then describe how this knowledge is leveraged to support adaptive small-cell selection.

\subsection{Knowledge Generation Module}
The Knowledge Generation module is responsible for converting raw monitoring data collected from the network into actionable control decisions. Within the proposed framework, knowledge corresponds to the optimal association decision that minimizes the objective function defined in Section~III.B while satisfying the system constraints.

At each decision epoch, the module evaluates the current network state, including traffic load, queue behavior, and service dynamics, and determines the most appropriate serving small cell for every user. These optimized association outcomes form structured knowledge entries that are subsequently stored in the knowledge base. Over time, this repository accumulates valuable operational experience, enabling the system to support both analytical optimization and learning-based approximation. To enable such knowledge extraction, the service behavior of each small cell must first be characterized through an appropriate analytical model.

\subsubsection{Small Cell Modeling} \label{smallcellmodel}

In dense 6G deployments, user performance is largely governed by congestion and scheduling contention within small cells. To capture these effects in a tractable yet analytically grounded manner, each small cell $i$ is modeled as an equivalent single-server queue, abstracting the aggregate impact of shared radio resources and scheduling dynamics rather than literal sequential service.

Let $\lambda_i(t)$ denote the aggregate arrival rate and $\mu_i(t)$ the effective service rate determined by radio resource availability, channel conditions, and scheduling efficiency. The arrival process is modeled as a renewal process with finite variance, while the service time follows a general distribution reflecting channel and processing variability. Assuming work-conserving operation at the aggregate level, each cell is approximated as a G/G/1 system. The stability condition is given in Eq.~\eqref{rho}, where the utilization factor must satisfy $0 \leq \rho_i(t) < 1$.

\small
\begin{equation}
\label{rho}
\rho_i(t) = \frac{\lambda_i(t)}{\mu_i(t)}, \quad 0 \leq \rho_i(t) < 1 
\end{equation}
\normalsize

Congestion is characterized through two analytical descriptors: the average queueing delay $W_i(t)$ and the average queue length $Q_i(t)$. As given in Eq.~\eqref{kingman}, the delay is approximated using Kingman’s formula for a G/G/1 queue.

\small
\begin{equation}
\label{kingman}
W_i(t) \approx 
\frac{\rho_i(t)}{1 - \rho_i(t)}
\cdot
\frac{c_{a,i}^2 + c_{s,i}^2}{2}
\cdot
\frac{1}{\mu_i(t)} ,
\end{equation}
\normalsize

Here, $c_{a,i}^2$ and $c_{s,i}^2$ denote the squared coefficients of variation of the inter-arrival and service processes, respectively. Applying Little’s law, the corresponding queue-length indicator is obtained as given in Eq.~\eqref{little}.

\small
\begin{equation}
\label{little}
Q_i(t) = \lambda_i(t) W_i(t) .
\end{equation}
\normalsize

The quantities $W_i(t)$ and $Q_i(t)$ therefore serve as analytically grounded congestion indicators within the optimization framework. This formulation avoids restrictive M/M/1 assumptions while preserving tractability and enabling heterogeneous traffic dynamics to be systematically incorporated into association decisions.

\subsubsection{Packet Service Degradation Indicator}

In addition to delay-related effects, heavy traffic within a small cell may also degrade packet-handling performance. In practical systems, congestion can cause buffer overflows, prolonged scheduling delays, or repeated retransmissions, all of which degrade the quality of packet delivery. To incorporate these effects into the association decision process, we introduce a packet-related service indicator that reflects the likelihood of packet service degradation under high load conditions.

As discussed in Section~IV.A, the service condition of each small cell is characterized by the queue-related descriptors $Q_i(t)$ and $W_i(t)$. When the queue length becomes large relative to the cell's effective service capacity, newly arriving packets are more likely to experience excessive waiting times or be discarded due to buffer limitations. Motivated by this observation, the packet service degradation indicator is defined as given in Eq. \ref{packetloss}.

\small
\begin{equation} 
\label{packetloss}
P_{ij}(t) = \beta_1 
\left(
\frac{Q_i(t)}{\tilde{C}_i(t)}
\right)^{\gamma}
\end{equation}
\normalsize

In this formulation, $\tilde{C}_i(t)$ denotes an equivalent service capacity descriptor for small cell $i$, expressed in the same units as the queue length indicator $Q_i(t)$. Conceptually, this term represents the effective buffering and packet handling capability of the cell under the current resource configuration. The ratio $\frac{Q_i(t)}{\tilde{C}_i(t)}$ therefore provides a normalized measure of congestion level. The exponent $\gamma \geq 1$ controls the indicator's sensitivity to congestion. Larger values of $\gamma$ emphasize high-load conditions and penalize heavily congested cells more aggressively. The parameter $\beta_1$ is a non-negative scaling coefficient that determines the relative contribution of this indicator within the overall objective function introduced in Eq.~\ref{target}.

Although the degradation level primarily depends on the state of small cell $i$, the notation $P_{ij}(t)$ is adopted to reflect that all users $j$ associated with that cell are affected by the same congestion condition. Under light load, when $Q_i(t) \ll \tilde{C}_i(t)$, the normalized occupancy remains small and the degradation indicator approaches zero. Through this formulation, $P_{ij}(t)$ functions as a load-aware penalty that discourages associations with heavily congested cells.

\subsubsection{Energy Service Cost Indicator}

Energy consumption affects both user experience and operational efficiency in dense 6G networks. In small-cell deployments, the energy required to maintain a communication link depends not only on channel conditions but also on the traffic state of the serving cell. When congestion increases, longer transmissions, scheduling delays, and retransmissions raise the overall communication energy cost.

Rather than modeling exact device battery depletion or detailed base-station power consumption, the proposed framework introduces an energy-related service indicator. This indicator captures the relative communication energy cost of maintaining a connection between user $j$ and small cell $i$ at time slot $t$, enabling energy considerations to be integrated into association decisions without device-specific power models. The baseline communication energy is mainly determined by the transmit and receive power levels, $P_{tx,ij}(t)$ and $P_{rx,ij}(t)$. To incorporate congestion effects, the queue-related indicator introduced in Section~IV.A is included. The resulting energy service cost indicator is defined as given in Eq. \ref{energyloss}.

\small
\begin{equation}
\label{energyloss}
E_{ij}(t) = \beta_2 \big( P_{tx,ij}(t) + P_{rx,ij}(t) \big)
\left( 1 + \delta \frac{Q_i(t)}{\tilde{C}_i(t)} \right)
\end{equation}
\normalsize

Here, $\tilde{C}_i(t)$ denotes the effective service capacity indicator and $Q_i(t)$ represents the queue-length descriptor of cell $i$, so $\frac{Q_i(t)}{\tilde{C}_i(t)}$ reflects the normalized congestion level. The parameter $\delta \geq 0$ controls how strongly congestion increases the energy cost: when $\delta = 0$, the model depends only on transmit and receive power, while larger values penalize associations with congested cells. The coefficient $\beta_2$ scales the impact of the energy indicator relative to the other service indicators in the objective function. Under light load ($Q_i(t) \approx 0$), the indicator reduces to $E_{ij}(t) \approx \beta_2 (P_{tx,ij}(t) + P_{rx,ij}(t))$, whereas under heavy traffic the congestion term dominates.

Thus, $E_{ij}(t)$ acts as a congestion-aware energy cost indicator within the optimization framework. Similar to the other indicators, it does not represent an exact physical energy measurement but provides a normalized metric that discourages energy-inefficient associations and contributes to minimizing the objective in Eq.~\ref{target}.

\subsubsection{Delay-Related Service Indicator}
Latency is one of the most critical performance metrics determining user-perceived quality of service in dense wireless networks. In small-cell deployments, delay is largely influenced by traffic congestion, scheduling contention, and queue buildup at the serving cell. For this reason, delay considerations are explicitly incorporated into the objective function defined in Eq.~\ref{target}.

Rather than attempting to estimate the exact end-to-end packet delay, which depends on multiple protocol layers and network segments, the proposed framework introduces a delay-related service indicator. This indicator captures how service conditions at a given small cell affect the likelihood of latency degradation for associated users.

As described in Section~IV.A, the operational state of each small cell $i$ at time slot $t$ is characterized by two queue-related descriptors: the queue-length indicator $Q_i(t)$ and the average waiting-time indicator $W_i(t)$. A large queue length signifies stronger contention for shared radio and scheduling resources, whereas a longer waiting time reflects increased buffering delay. Since both effects contribute to service latency, they are jointly incorporated into the delay-related indicator. Accordingly, the delay-related service indicator for user $j$ associated with small cell $i$ at time slot $t$ is defined as given in Eq. \ref{timeloss}.

\small
\begin{equation}
\label{timeloss}
D_{ij}(t) = 
\eta_1 \frac{Q_i(t)}{\tilde{C}_i(t)}
+
\eta_2 \mu_i(t) W_i(t)
\end{equation}
\normalsize

In this formulation, $\tilde{C}_i(t)$ denotes the effective service capacity descriptor introduced earlier. Since $\tilde{C}_i(t)$ is expressed in the same unit as $Q_i(t)$, the ratio $\frac{Q_i(t)}{\tilde{C}_i(t)}$ provides a normalized and dimensionless measure of congestion level. The second term captures delay tendencies through the average waiting time. The quantity $W_i(t)$ has units of time, while $\mu_i(t)$ represents the effective service rate (in $\text{s}^{-1}$). Consequently, the product $\mu_i(t) W_i(t)$ is dimensionless and reflects the relative waiting intensity experienced within the cell. The parameters $\eta_1 \geq 0$ and $\eta_2 \geq 0$ are weighting coefficients that regulate the relative contribution of congestion level and waiting-time effects. Through this formulation, the overall delay indicator $D_{ij}(t)$ remains dimensionless, ensuring consistency when combined with the packet-related and energy-related indicators in the objective function. When the cell operates under light load conditions ($Q_i(t) \ll \tilde{C}_i(t)$ and $W_i(t)$ is small), both terms remain limited and the delay indicator takes a low value. As congestion increases, the normalized queue occupancy and waiting-time intensity grow simultaneously, causing the indicator to rise and signaling a higher risk of latency degradation.

Therefore, $D_{ij}(t)$ functions as a congestion-aware delay descriptor within the proposed optimization framework. Similar to the other service indicators, it is not intended to represent an exact physical delay value but rather a normalized metric that discourages associations likely to result in higher latency. This indicator directly contributes to the minimization of the objective function defined in Eq.~\ref{target}.

\normalsize
\subsubsection{Lagrangian-Based Optimization for Objective Function}

The small-cell selection problem defined in Eq.~\ref{target}, subject to the constraints in Eqs.~\ref{c1} and \ref{c2}, is a binary integer optimization problem. Because of the discrete nature of $X_{ij}(t)$, the formulation is combinatorial and difficult to solve directly with gradient-based techniques. To obtain a tractable representation, a continuous relaxation is applied where $X_{ij}(t) \in [0,1]$, interpreted as fractional association weights, and the binary constraint is later enforced through projection. Under this relaxation, the Lagrangian function for time slot $t$ is constructed as shown in Eq.~\ref{lagrange}. The multipliers $\nu_j(t)$ and $\xi_i(t)$ correspond to the user-association and cell-capacity constraints, respectively.

\small
\begin{equation} \label{lagrange}
\begin{aligned}
\mathcal{L}(X,\nu,\xi) &=
\sum_{j=1}^{M} \sum_{i=1}^{N}
\left( D_{ij}(t) + P_{ij}(t) + \alpha E_{ij}(t) \right) X_{ij}(t) \\
&+ \sum_{j=1}^{M} \nu_j(t) \left( \sum_{i=1}^{N} X_{ij}(t) - 1 \right) \\
&+ \sum_{i=1}^{N} \xi_i(t) \left( \sum_{j=1}^{M} X_{ij}(t) - C_i(t) \right).
\end{aligned}
\end{equation}
\normalsize

Applying the stationarity condition to Eq.~\ref{lagrange}, the derivative with respect to $X_{ij}(t)$ yields the gradient in Eq.~\ref{solution}.

\small
\begin{equation}\label{solution}
\frac{\partial \mathcal{L}}{\partial X_{ij}(t)}
=
D_{ij}(t) + P_{ij}(t) + \alpha E_{ij}(t) + \nu_j(t) + \xi_i(t)
\end{equation}
\normalsize

This result shows that assignments with high service cost or strong dual penalties become less favorable. The multipliers $\nu_j(t)$ and $\xi_i(t)$ are updated using a projected subgradient rule to enforce the constraints. After solving the relaxed problem, a feasible binary association is obtained through projection. Each user $j$ is assigned to the cell minimizing the adjusted cost in Eq.~\ref{adjust} while respecting cell capacity.

\small
\begin{equation}\label{adjust}
i^*(j,t) = \arg\min_{i} \left( D_{ij}(t) + P_{ij}(t) + \alpha E_{ij}(t) + \xi_i(t) \right)
\end{equation}
\normalsize

The Lagrangian framework therefore balances user-level service indicators with system-level capacity constraints. However, repeatedly solving the optimization under dynamic network conditions may introduce computational overhead. For this reason, the resulting optimal associations are later used as supervisory signals for training a machine-learning module that approximates these decisions with significantly lower runtime complexity.

\vspace{-0.15in}
\normalsize
\subsection{Machine Learning Module}

Although the Lagrangian-based optimization framework provides a principled solution for the small-cell selection problem, solving the optimization repeatedly at every time slot may introduce noticeable computational overhead in dense 6G deployments. As the number of users and candidate cells increases, recomputing association decisions can limit real-time responsiveness. To mitigate this issue, a lightweight machine learning module is introduced to approximate the decisions produced by the optimization engine.

Within the proposed KDN architecture, the learning module operates in the Knowledge Plane and complements the optimization framework rather than replacing it. The Lagrangian optimization acts as a knowledge generator that produces high-quality association decisions under varying network conditions. These decisions are stored in the knowledge base and used as labeled samples to train a predictive model. Once trained, the model can emulate the optimization behavior with significantly lower computational cost, enabling faster decisions during network operation.

Each training instance corresponds to a candidate user–cell pair $(i,j)$ observed at time slot $t$. For every pair, a feature vector is constructed using the service indicators defined earlier. The feature representation is defined as $x_{ij}(t) =
[
Q_i(t)/\tilde{C}_i(t),\;
\mu_i(t) W_i(t),\;
P_{ij}(t),\;
D_{ij}(t),\;
E_{ij}(t)]$. All elements represent dimensionless service indicators. The normalized queue ratio and waiting-time term describe congestion conditions, while the packet, delay, and energy indicators reflect service costs appearing in the optimization objective. Using these indicators directly enables the learning model to internalize the decision structure of the optimization framework. To ensure generalization across users, explicit user identifiers are intentionally excluded. The class label $y_{ij}(t)$ corresponds to the small cell selected by the optimization module.

\begin{algorithm}[h]
\caption{LVQ-Based Small Cell Selection}
\small
\label{algo}
\begin{algorithmic}[1]

\Require Training dataset $(x_n, y_n)$ generated by Lagrangian optimization
\Require Diminishing learning rate schedule $\{\alpha^{(e)}\}_{e=1}^{E}$ such that $\alpha^{(e+1)} < \alpha^{(e)}$ and $\lim_{e\to\infty}\alpha^{(e)} = 0$
\Ensure Prototype set $W = \bigcup_{i=1}^{N_s} W_i$

\State \textbf{Prototype Initialization}
\For{each small cell $i$}
    \State Initialize prototype vectors $w \in W_i$ using training samples with label $i$
\EndFor

\State \textbf{Training Phase}
\For{epoch $e = 1$ to $E$}
    \For{each training sample $(x_n, y_n)$}
        \State Identify closest prototype 
        \State $w_* = \arg\min_{w \in W} \|x_n - w\|$
        \If{$\text{label}(w_*) = y_n$}
            \State $w_* \leftarrow w_* + \alpha^{(e)}(x_n - w_*)$
        \Else
            \State $w_* \leftarrow w_* - \alpha^{(e)}(x_n - w_*)$
        \EndIf
    \EndFor
    \State Update learning rate $\alpha^{(e+1)} < \alpha^{(e)}$
\EndFor

\State \textbf{Online Small Cell Selection}
\For{each user $j$ at time slot $t$}
    \For{each candidate small cell $i$}
        \State Compute feature vector $x_{ij}(t)$
        \State $d_i(t) = \min_{w \in W_i} \|x_{ij}(t) - w\|$
    \EndFor
    \State Select serving cell $i^*(j,t) = \arg\min_i d_i(t)$
\EndFor

\end{algorithmic}
\end{algorithm}

To approximate the association decisions, Learning Vector Quantization (LVQ) is adopted due to its simplicity and low computational cost. In LVQ, each class represents a candidate small cell and is modeled by one or more prototype vectors in the feature space. These prototypes capture typical service conditions under which a cell is selected by the optimization framework. LVQ is particularly suitable for dense 6G edge environments because inference requires only simple distance calculations, the model maintains a compact set of prototypes, and incremental updates can be performed as new optimization outcomes become available.

\normalsize
During training, prototype vectors are iteratively updated using labeled samples generated by the optimization module. Let $x_n$ denote a feature vector and $y_n$ its corresponding small-cell label. The closest prototype in Euclidean distance is determined as given in Eq. \ref{dist}.

\small
\begin{equation}
w_* = \arg\min_{w \in W} \|x_n - w\|.
\label{dist}
\end{equation}
\normalsize

Based on this, the update rule is given by Eq. \ref{move}.
\small

\begin{equation}
w_* \leftarrow
\begin{cases}
w_* + \alpha^{(e)}(x_n - w_*), & \text{if } \text{label}(w_*) = y_n, \\
w_* - \alpha^{(e)}(x_n - w_*), & \text{otherwise}.
\end{cases}
\label{move}
\end{equation}
\normalsize

Once training is completed, the LVQ model can be used for real-time association. For each user $j$ at time slot $t$, the serving cell is selected as given in Eq. \ref{dit2}.

\small
\begin{equation}
d_i(t) = \min_{w \in W_i} \|x_{ij}(t) - w\|,
\label{dit}
\end{equation}
\begin{equation}
i^*(j,t) = \arg\min_i d_i(t).
\label{dit2}
\end{equation}
\normalsize

This inference mechanism avoids repeatedly solving the full optimization problem while preserving the decision structure learned from the Lagrangian framework. Periodically, the optimization module can be reactivated to generate new labeled samples reflecting updated traffic patterns or network conditions. These samples are then used to refine the LVQ prototypes and maintain model accuracy.

From a computational perspective, the original optimization involves solving a constrained assignment problem whose complexity grows with the number of users and cells. In contrast, the LVQ inference stage requires only a limited number of prototype-distance evaluations. Therefore, the proposed hybrid optimization–learning approach substantially reduces runtime complexity while maintaining high-quality association decisions.

\subsection{Knowledge Storage Module}

Complementing the knowledge generation and learning components, the proposed architecture includes a dedicated knowledge storage module that preserves historical network intelligence, as illustrated in Fig.~\ref{fig3}. Unlike conventional repositories that mainly archive raw monitoring data, this module stores structured information derived from the optimization engine and the learning process. The stored records include time-stamped user–cell association decisions, the corresponding service indicators, and the observed network performance after each decision. As a result, the repository captures not only instantaneous network states but also the context behind control actions. This historical knowledge plays a key role in the KDN control loop. First, it provides labeled datasets required for training and periodically updating machine learning models in the knowledge plane. As the network evolves, newly generated optimization outcomes are continuously added to the repository, enabling incremental refinement of the learned association policies. Second, the stored information supports longitudinal analysis of network behavior, including performance evaluation, anomaly detection, and policy adjustment. By maintaining a structured history of states and decisions, the system can identify recurring patterns in traffic demand, congestion dynamics, and energy-related service conditions common in dense 6G deployments. Furthermore, the knowledge storage module reinforces the closed-loop nature of the proposed KDN architecture. Decisions applied in the data plane can be traced back to their originating indicators and optimization outcomes, improving transparency and enabling system-level diagnostics. In this way, the module acts as the persistent memory of the architecture, supporting continuous learning, adaptive optimization, and knowledge-driven control within the proposed KDN-based small-cell selection framework.

\vspace{-0.19in}
\section{Performance Evaluation}
This section presents the simulation framework and evaluation methodology used to assess the proposed KDN-enabled small-cell selection approach. The goal is to quantify how the Lagrangian–learning framework improves service quality, congestion management, and overall network efficiency in dense small-cell deployments. The simulations are implemented using the NS-3 network simulator with the LTE/5G module for cellular network modeling. The proportional fair scheduler provided by NS-3 is used at the MAC layer for resource allocation. Channel propagation follows the log-distance path-loss model, combined with log-normal shadowing and Rayleigh small-scale fading. Traffic generation is implemented using the built-in UDP CBR application model. Packet queues are maintained at the base-station MAC layer, where buffer overflow results in packet drops under the DropTail policy. These configuration choices allow the simulator to capture realistic interactions between radio conditions, traffic dynamics, and scheduling behavior in dense cellular environments.

\begin{table}[h]
\centering
\caption{Topology, Mobility, and Traffic Parameters}
\label{tab:networkparams}
\scriptsize
\renewcommand{\arraystretch}{1}
\begin{tabular}{ll}
\toprule
\textbf{Parameter} & \textbf{Value} \\
\midrule
Simulation Area & $1000 \times 1000$ m$^2$ \\
Number of Macro Base Stations & 1 \\
Number of Small Cells ($N_s$) & 10 -- 30 (default: 16) \\
Inter-site Distance (avg.) & 150 -- 250 m \\
Small Cell Deployment & Uniform Random \\
Minimum SBS Separation & 40 m \\

Number of Users ($M$) & 40 -- 160 (default: 80) \\
User Density & 40 -- 160 users/km$^2$ \\
User Initial Distribution & Uniform Random \\

User Mobility Model & Random Walk \\
User Speed Range & 0.5 -- 1.5 m/s \\
Mobility Update Interval & 1 s \\
Boundary Handling & Reflection at edges \\

Traffic Type & UDP Constant Bit Rate (CBR) \\
Packet Size & 512 bytes \\
Traffic Rate per User & 200 kbps \\
Number of Active Flows per User & 1 \\
Flow Direction & Uplink + Downlink \\

Queue Type at SBS & Per-user logical queues \\
MAC Buffer Size & 1000 packets \\
Packet Drop Policy & DropTail \\

Warm-up Period & 60 s \\
Random Seed Control & Independent per run \\
\bottomrule
\end{tabular}
\end{table}
\vspace{-0.1in}
\subsection{Simulation Details}
This subsection describes the simulation environment with network topology, traffic model, mobility assumptions, and radio parameters. The overall configuration is chosen to represent a dense heterogeneous cellular deployment in which load imbalance, interference, and queue dynamics play a critical role in user association decisions.

\subsubsection{Network Topology}
A two-tier heterogeneous cellular network is deployed over a $1000 \times 1000$ m$^2$ area. A macro base station (MBS) is placed at the center to provide umbrella coverage, while $N_s$ small-cell base stations (SBSs) are distributed across the region to increase spatial reuse and support edge users. To analyze performance under varying network densities, the number of small cells and users is varied across simulation scenarios, with $10 \leq N_s \leq 30$ and $40 \leq M \leq 160$. Unless otherwise stated, the default configuration uses $N_s = 16$ and $M = 80$. Both SBSs and users are initialized using uniform spatial distributions, producing heterogeneous radio conditions and non-uniform traffic loads typical of dense urban deployments. Depending on the SBS density, the average inter-site distance ranges approximately between 150 m and 250 m, enabling the evaluation of association strategies under different congestion levels.

\subsubsection{Mobility Model}
All base stations remain fixed during the simulation. Mobile users follow the Random Walk Mobility Model with bounded velocity to emulate pedestrian movement in urban environments. User speeds range between $0.5$ and $1.5$ m/s, and positions are updated every second. When a user reaches the simulation boundary, a reflection mechanism keeps the node within the area. This configuration generates time-varying channel conditions and dynamic association opportunities without introducing unrealistic mobility patterns.

\begin{table}[h]
\centering
\caption{Radio and PHY/MAC Parameters}
\label{tab:phyparams}
\scriptsize
\renewcommand{\arraystretch}{1}
\begin{tabular}{p{4.2cm}  p{3.8cm}}
\toprule
\textbf{Parameter} & \textbf{Value} \\
\midrule
Carrier Frequency & 3.5 GHz \\
System Bandwidth & 20 MHz \\
Transmission Time Interval (TTI) & 1 ms \\

BS Transmission Power (SBS) & 30 dBm \\
BS Transmission Power (MBS) & 43 dBm \\
UE Maximum Transmission Power & 23 dBm \\

Antenna Configuration (BS) & $4 \times 4$ MIMO \\
Antenna Configuration (UE) & $2 \times 2$ MIMO \\

Path Loss Model & Log-Distance + Log-Normal Shadowing \\
Shadowing Std. Dev. & 8 dB \\
Small-Scale Fading & Rayleigh \\

Thermal Noise Density & $-174$ dBm/Hz \\
System Bandwidth for Noise Calc. & 20 MHz \\
Receiver Noise Figure (BS) & 5 dB \\
Receiver Noise Figure (UE) & 7 dB \\
Approx. Noise Power (BS) & $\approx -96$ dBm \\
Approx. Noise Power (UE) & $\approx -94$ dBm \\

MAC Scheduler & Proportional Fair \\

\bottomrule
\end{tabular}
\end{table}

\subsubsection{Traffic Model}
Each user continuously generates both uplink and downlink traffic using a UDP-based Constant Bit Rate (CBR) model. Packets of 512 bytes are transmitted at 200 kbps per direction, creating sustained load across the network. This setup induces queue buildup and resource contention at SBs, enabling realistic congestion dynamics. Packets are buffered at the serving base station whenever radio resources are unavailable, and the resulting queue statistics are later used to compute the service indicators of the proposed framework.

\subsubsection{Radio and Channel Configuration}
All base stations operate at 3.5 GHz, representing a typical mid-band deployment for emerging 5G-Advanced and early 6G systems. The system bandwidth is 20 MHz, providing a practical trade-off between spectral efficiency and inter-cell interference in dense networks. Base stations employ $4 \times 4$ MIMO while user devices use $2 \times 2$ MIMO. Propagation is modeled using a log-distance path loss model with log-normal shadowing to capture large-scale channel variations, while small-scale fading follows a Rayleigh model typical of non-line-of-sight urban environments. Because all cells share the same carrier frequency, inter-cell interference naturally arises. Radio resources are scheduled using a proportional fair policy that balances throughput and fairness under varying channel conditions. These settings capture both radio-layer dynamics and traffic-driven congestion, enabling realistic system states from which the queue indicators $Q_i(t)$ and $W_i(t)$ are obtained for evaluating the proposed framework.

\subsubsection{Queueing and Service Modeling}
Each small cell maintains separate logical queues for user traffic at the MAC layer. Resource allocation follows a proportional fair scheduler, causing the effective service rate $\mu_i(t)$ to vary with channel conditions and user load, reflecting practical cellular scheduling behavior. Queue statistics are collected at every decision interval, including queue length $Q_i(t)$ and average waiting time $W_i(t)$. These metrics characterize cell congestion and are used to compute the service indicators $P_{ij}(t)$, $D_{ij}(t)$, and $E_{ij}(t)$ defined earlier. The resulting indicators form the input to the Lagrangian optimization module and serve as feature vectors for the LVQ learning model, keeping the simulation environment and learning framework tightly coupled.

\begin{table}[h]
\centering
\caption{Optimization and Learning Parameters}
\label{tab:mlparams}
\scriptsize
\renewcommand{\arraystretch}{1}
\begin{tabular}{p{4.2cm}  p{3.8cm}}
\toprule
\textbf{Parameter} & \textbf{Value} \\
\midrule
Decision Interval ($T_d$) & 1 s \\
Total Simulation Time ($T_{\text{sim}}$) & 600 s \\
Number of Independent Runs & 20 \\
Confidence Interval & 95\% \\

Optimization Solver & Sequential Quadratic Programming \\
Max Optimization Iterations & 200 \\
Optimization Re-trigger Period & Every 20 intervals \\

Initial Supervised Data Collection Phase & First 50 intervals \\

LVQ Variant & LVQ2.1 \\
Number of Prototypes per Cell & 2 \\
Initial Learning Rate & 0.1 \\
Learning Rate Decay & Exponential decay \\

Number of Input Features & 6 \\
Feature Normalization & Min–Max Scaling \\
Distance Metric in LVQ & Euclidean Distance \\

Label Source & Lagrangian Optimization Output \\
Model Update Mode & Online Incremental Update \\

\bottomrule
\end{tabular}
\end{table}

\subsubsection{Optimization and Learning Operation}
The simulation operates in decision intervals of $T_d = 1$ s. At each interval, the knowledge plane gathers monitoring data and computes service indicators for all user–cell pairs. During the first 50 intervals, the Lagrangian optimization module runs at every step to generate reference association decisions, which are applied to the network and stored as labeled samples for training the LVQ model. After this bootstrapping phase, the system switches to a hybrid mode where the LVQ model predicts associations in most intervals, significantly reducing computational overhead. To maintain accuracy under changing traffic and mobility conditions, the optimization module is reactivated every 20 intervals, and the new optimal decisions are used to incrementally update the LVQ prototypes. This hybrid strategy preserves optimization accuracy while enabling real-time responsiveness in dense networks.

\subsubsection{Simulation Scale and Duration}
Each simulation run lasts $T_{\text{sim}} = 600$ s, allowing the network to experience multiple mobility events, traffic fluctuations, and load imbalance conditions across cells. A warm-up period of the first 60 s is excluded from evaluation to eliminate initialization bias related to queue buildup and initial association states. All performance metrics are averaged over 20 independent simulation runs with distinct random seeds controlling user placement, mobility trajectories, traffic generation, and channel fading realizations. For fair comparison, the same set of random seeds is applied across all evaluated schemes in corresponding runs, enabling paired statistical analysis and reducing variance induced by stochastic factors. For each metric, 95\% confidence intervals are computed using the Student's $t$-distribution as $\bar{x} \pm t_{0.975,n-1} \frac{s}{\sqrt{n}}$, where $\bar{x}$ denotes the sample mean, $s$ is the sample standard deviation, and $n=20$ is the number of independent runs. The relative standard deviation remains limited (typically below 5\% depending on the metric), indicating stable behavior across stochastic realizations. Moreover, the confidence intervals of the proposed framework remain non-overlapping with those of baseline schemes under moderate and high load conditions, confirming the statistical significance of the reported gains at the 95\% confidence level.

\begin{figure*}[h]
\centering
\includegraphics[width=0.76\textwidth]{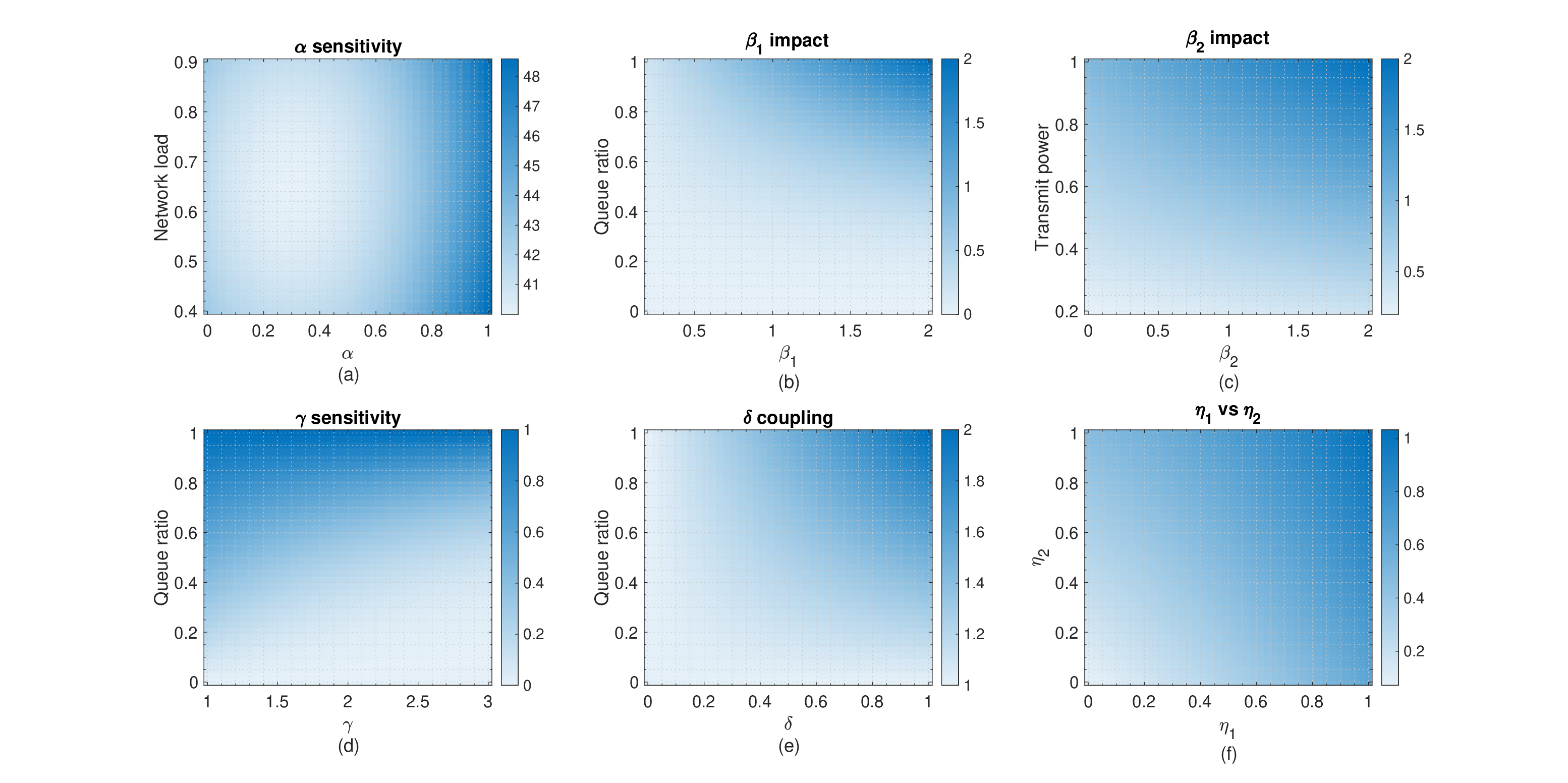}
\caption{Parameter calibration and stability analysis.}
%\vspace{-2em}
\label{param}
\end{figure*}

\subsubsection{Dataset Generation and Training Procedure}
Training data for the LVQ model is generated online during the simulation through the optimization module. During the initial bootstrapping phase, the Lagrangian optimization algorithm determines the optimal association decisions for all users at every decision interval. Each user–cell candidate pair together with its service indicators forms a labeled training sample. In the default configuration, approximately $M \times N_s$ candidate samples are produced at each decision interval. For example, with $M=80$ users and $N_s=16$ small cells, roughly $1280$ labeled samples are generated per interval. Over the initial 50 training intervals, this results in approximately $6.4 \times 10^4$ training instances. To avoid overfitting and ensure stable learning behavior, the collected dataset is randomly divided into training and validation subsets using an 80/20 split. The training subset is used to update LVQ prototypes, while the validation subset monitors classification consistency relative to the optimization decisions. After the initial training stage, the model operates in an online learning mode. Every 20 decision intervals, the optimization module is invoked again to generate fresh labeled samples reflecting updated traffic and mobility conditions. These samples are used to incrementally refine the LVQ prototypes, allowing the learning model to adapt to long-term changes in the network environment.

\subsubsection{Knowledge Graph Construction and Update}
In the proposed KDN framework, network observations and decision outcomes are captured via a lightweight knowledge graph representing users, small cells, traffic conditions, and service indicators. The graph comprises three entity types—users, small cells, and network states—linked by edges encoding association decisions, observed service metrics, and performance outcomes. At each decision interval, the monitoring module records queue statistics, delay, packet degradation, and energy metrics for all user–cell pairs, storing them as entity-linked attributes. For a system with $M=80$ users and $N_s=16$ small cells, the graph contains roughly $96$ nodes and several hundred relational edges per interval, reflecting possible associations and observed states. The structure evolves incrementally: new network states and optimization results are appended, while outdated data is pruned via a sliding observation window. This dynamic update maintains a compact, informative representation of the network environment, supporting both optimization oversight and learning-based inference.

\subsubsection{Online--Offline Update Policy}
The hybrid optimization–learning framework operates over two time scales. The LVQ classifier performs association decisions at every scheduling interval, while the Lagrangian optimization module is executed periodically to refresh supervisory labels and update the prototype set. Unless otherwise stated, global re-optimization is triggered every $T_r = 200$ decision intervals, representing slow-timescale network evolution. This value is selected to balance computational overhead and QoS stability. To emulate realistic traffic variations, a deviation-based trigger is also implemented. Let $\Delta_{\text{QoS}}(t)$ denote the relative change in the composite service metric compared to its recent moving average. If $\Delta_{\text{QoS}}(t) > \delta$ (with $\delta = 5\%$ in simulations), early re-optimization is activated before the nominal $T_r$ period. During each update event, new optimization-generated labels are produced and the LVQ model is retrained. Since retraining complexity scales as $\mathcal{O}(N K d)$ and update events occur sparsely relative to the association interval, the amortized computational overhead remains significantly lower than continuous optimization.

\begin{table*}[h]
\centering
\caption{Comparison of Association Strategies and Design Characteristics}
\label{tab:baseline_features}
\scriptsize
\renewcommand{\arraystretch}{1}
\begin{tabular}{lcccccc}
\toprule
Method 
& Radio Awareness 
& Load Awareness 
& Queue Awareness
& QoS Awareness 
& Learning 
& Computational Cost \\
\midrule

Max-RSRP 
& \checkmark
& --
& --
& --
& --
& Very Low \\

Load-Aware Heuristic (LHR) 
& \checkmark
& \checkmark
& --
& Partial
& --
& Low \\

Greedy Service-Aware (GSA)
& \checkmark
& \checkmark
& \checkmark
& \checkmark
& --
& Medium \\

Proposed KDN Framework
& \checkmark
& \checkmark
& \checkmark
& \checkmark
& \checkmark
& Medium \\

\bottomrule
\end{tabular}
\end{table*}

\subsection{Parameter Calibration and Sensitivity Analysis}
The proposed framework includes several parameters controlling how delay, packet service degradation, and energy consumption affect the association decision. Although these parameters are introduced in Section IV, their values must be calibrated to ensure stable system behavior. Instead of heuristic selection, we perform a sensitivity analysis using the simulation setup described in Section V.A. Each parameter is varied within a feasible range while monitoring average latency and packet loss. The resulting trends are summarized through the heatmaps in Fig.~\ref{param}, from which stable operating regions are identified and the final parameter configuration is selected.

\subsubsection{Energy-Related Parameters}

Energy awareness is mainly governed by the parameters $\alpha$, $\beta_2$, and $\delta$. The coefficient $\alpha$ determines the relative importance of the energy indicator in the objective function, $\beta_2$ scales the energy service cost defined in Eq.~(\ref{energyloss}), and $\delta$ controls the influence of congestion on energy expenditure. Fig.~\ref{param}a shows the impact of varying $\alpha$ within $[0,1]$. Very small values prioritize delay and reliability, whereas large values bias the system toward energy-efficient cells even if delay increases. A stable region appears around $\alpha \approx 0.3$. The parameter $\beta_2$ is evaluated under different transmit power conditions (Fig.~\ref{param}c), where $\beta_2 = 0.5$ provides balanced influence. The congestion–energy coupling parameter $\delta$ (Fig.~\ref{param}e) increases the energy penalty in loaded cells; the analysis indicates that $\delta = 0.5$ captures this effect without exaggeration.

\subsubsection{Packet Service Parameters}
The packet service degradation indicator in Eq.~(\ref{packetloss}) depends on $\beta_1$ and $\gamma$. The parameter $\beta_1$ controls the contribution of packet degradation to the objective, while $\gamma$ determines the sensitivity to congestion. As illustrated in Fig.~\ref{param}b, very small $\beta_1$ values insufficiently penalize congested cells, whereas large values may cause excessive switching. The analysis indicates that $\beta_1 = 1$ offers a suitable balance. The exponent $\gamma$ controls how sharply degradation increases as queues approach capacity. Fig.~\ref{param}d shows that $\gamma = 2$ effectively reflects congestion while maintaining stability.
\subsubsection{Delay-Related Parameters}
The delay indicator in Eq.~(\ref{timeloss}) includes the weights $\eta_1$ and $\eta_2$, representing the influence of queue buildup and waiting time. Since both factors jointly determine service latency, appropriate weighting is necessary. By jointly varying these parameters (Fig.~\ref{param}f), the results show that slightly prioritizing queue buildup improves stability in dense areas. So, $\eta_1 = 0.6$ and $\eta_2 = 0.4$ are adopted. %Based on these, the final parameter configuration used in the simulations is summarized in Table~\ref{tab:parameter_values}.

%\item \textit{Average Energy Consumption:}
%This metric captures the average transmission energy consumed by small cells while serving users. It is computed as $\bar{E} = \frac{1}{T} \sum_{t=1}^{T} \sum_{i=1}^{N_s} E_i(t)$, where $E_i(t)$ denotes the energy consumed by small cell $i$ during time interval $t$, and $N_s$ is the total nsumber of small cells.

\vspace{-0.1in}
\subsection{Baseline Association Schemes}

The proposed optimization–learning framework is compared with three representative association strategies reflecting increasing levels of network awareness. Rather than reproducing all literature variants with heterogeneous assumptions, we select baselines that capture common design philosophies while ensuring fair comparison under identical topology, traffic, mobility, and radio conditions.

\begin{itemize}

\item {Maximum Signal Strength (Max-RSRP):} Users associate with the cell providing the strongest received signal power. This radio-centric method ignores load, delay, reliability, and energy factors, often resulting in imbalance in dense deployments.

\item {Load-Aware Heuristic Association (LHR):} Signal strength is combined with estimated cell load (e.g., queue occupancy or resource utilization) to improve distribution. However, the approach remains heuristic and does not explicitly model delay, packet degradation, or energy cost.

\item {Greedy Service-Aware Association (GSA):} Multiple service indicators, including delay tendency, packet degradation, and communication energy, are considered at each interval. Decisions rely on instantaneous conditions and may therefore be locally efficient but globally suboptimal.

\end{itemize}

Pure machine-learning baselines are excluded, as the objective is to assess whether a lightweight model can approximate optimization-driven decisions rather than replace them. A standalone neural model trained purely on historical data would optimize empirical patterns without explicitly enforcing the constrained multi-objective structure embedded in the proposed formulation, making the comparison methodologically inconsistent with the problem definition. The proposed framework formulates association as a constrained optimization problem jointly addressing delay, reliability, and energy efficiency, and employs LVQ to approximate these decisions for low-latency inference, with periodic optimization updates.

%\begin{table}[h]
%\centering
%\caption{Calibrated Model Parameters}
%\label{tab:parameter_values}
%\scriptsize
%\renewcommand{\arraystretch}{1}
%\begin{tabular}{lc}
%\toprule
%Parameter & Selected Value \\
%%\midrule
%Energy weight $\alpha$ & 0.3 \\
%Packet scaling $\beta_1$ & 1.0 \\
%Energy scaling $\beta_2$ & 0.5 \\
%Congestion exponent $\gamma$ & 2 \\
%Energy–congestion coupling $\delta$ & 0.5 \\
%Queue weight $\eta_1$ & 0.6 \\
%Waiting-time weight $\eta_2$ & 0.4 \\
%\bottomrule
%\end{tabular}
%\end{table}
\begin{figure*}[h]
\centering
\subfloat[]{%
\includegraphics[width=0.31\textwidth]{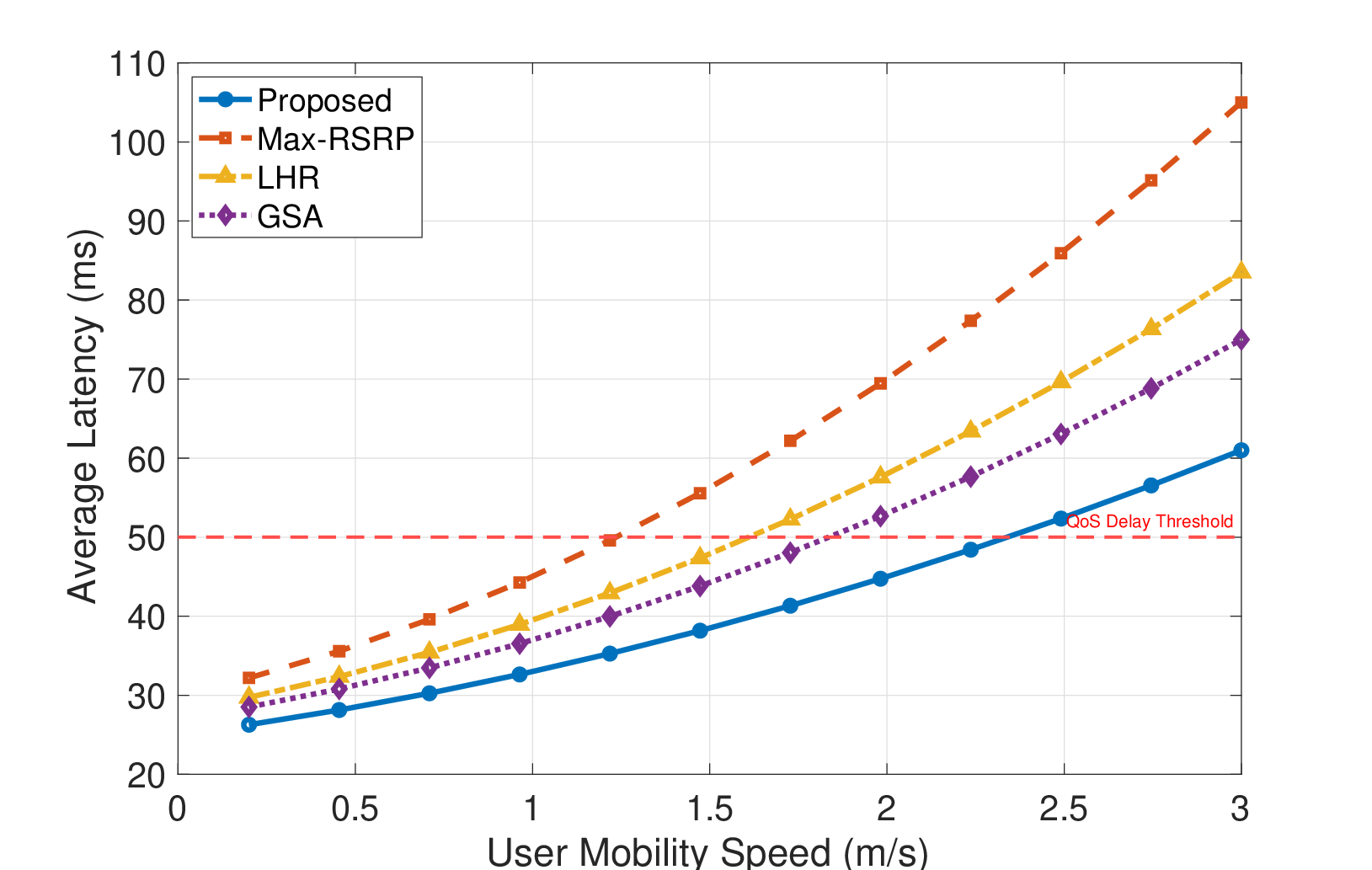}%
\label{f1}%
} %
\subfloat[]{%
\includegraphics[width=0.31\textwidth]{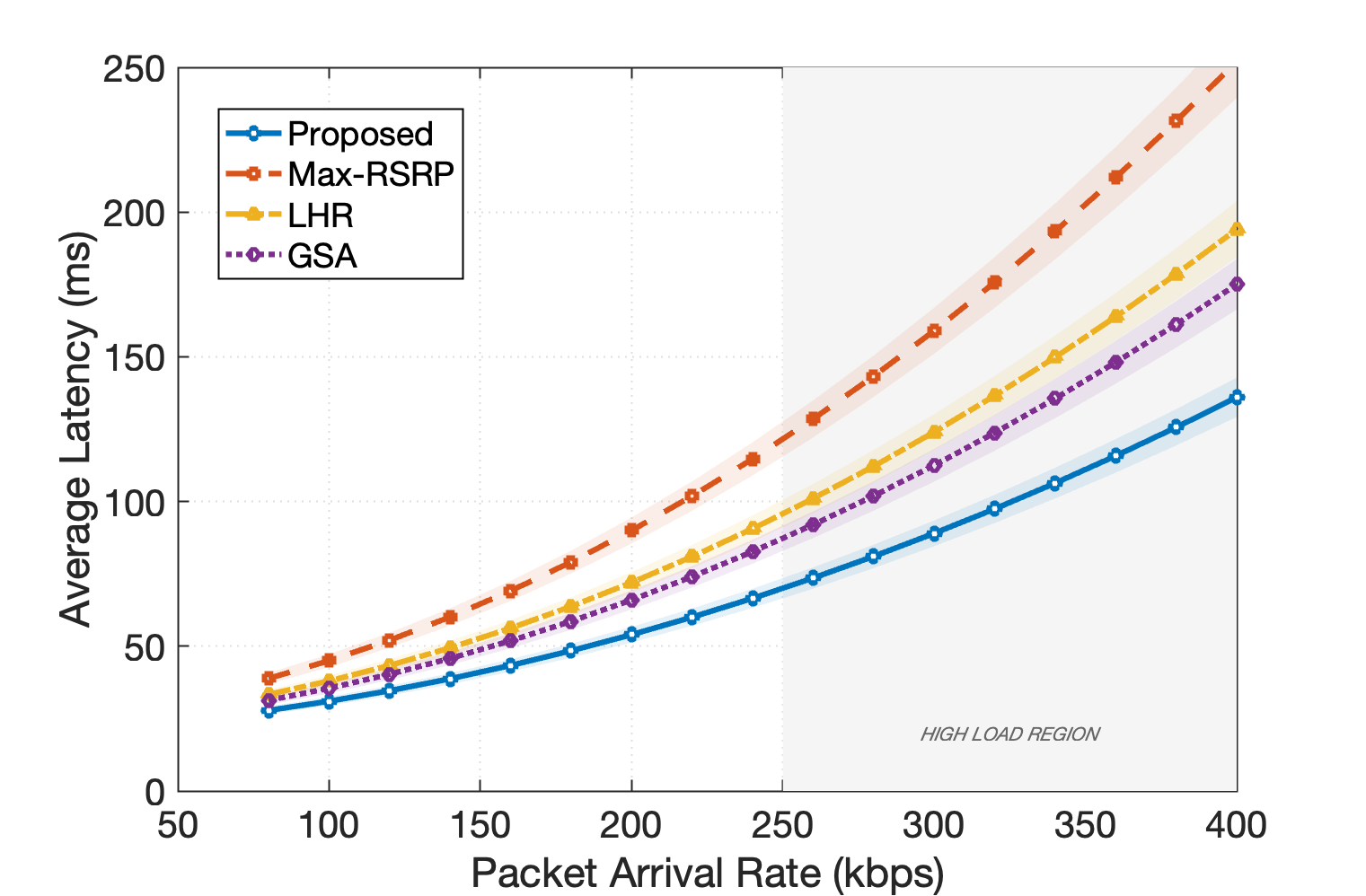}%
\label{f3}%
} %
\subfloat[]{%
\includegraphics[width=0.31\textwidth]{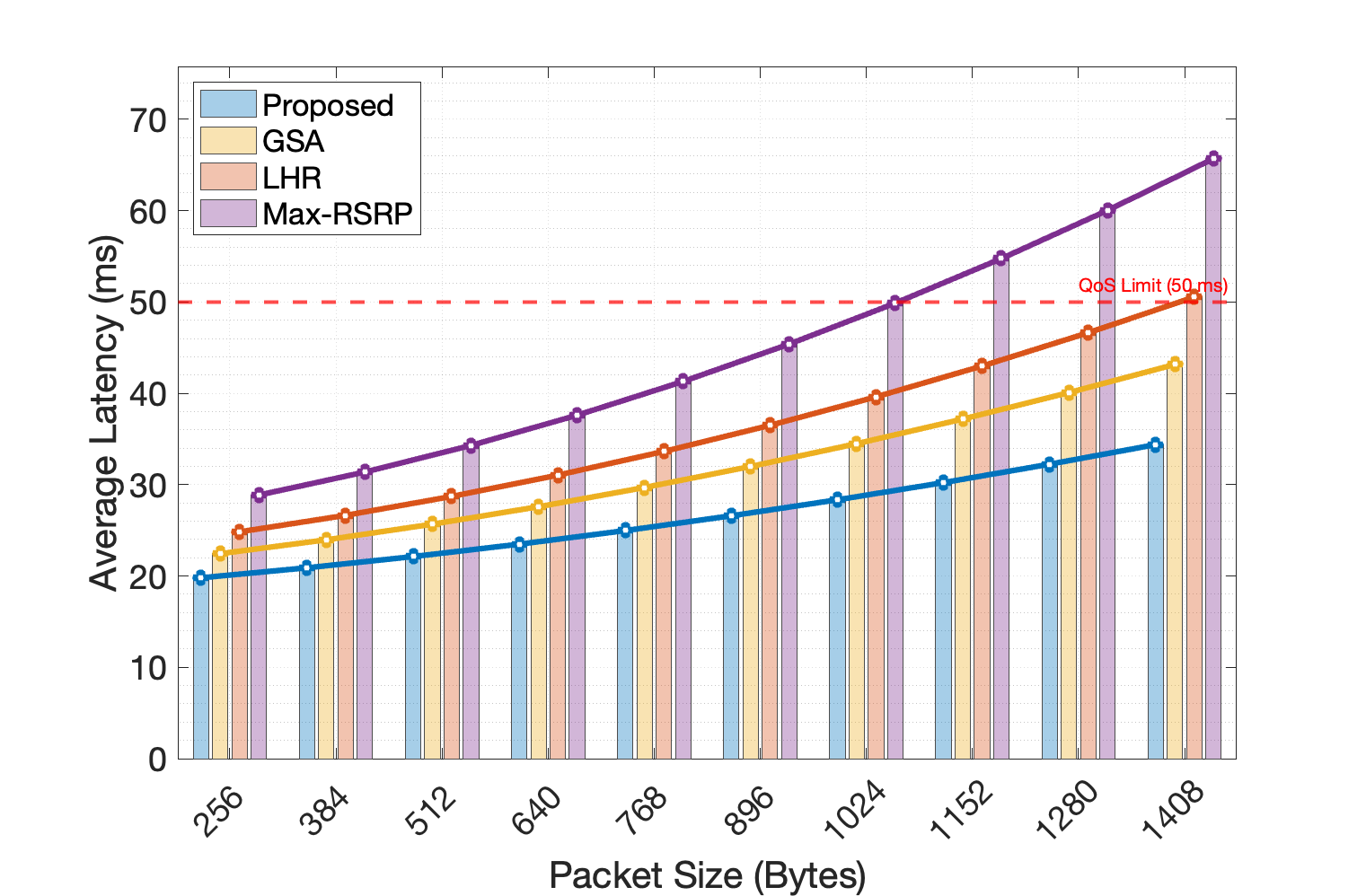}%
\label{f5}%
} \\
\subfloat[]{%
\includegraphics[width=0.48\textwidth]{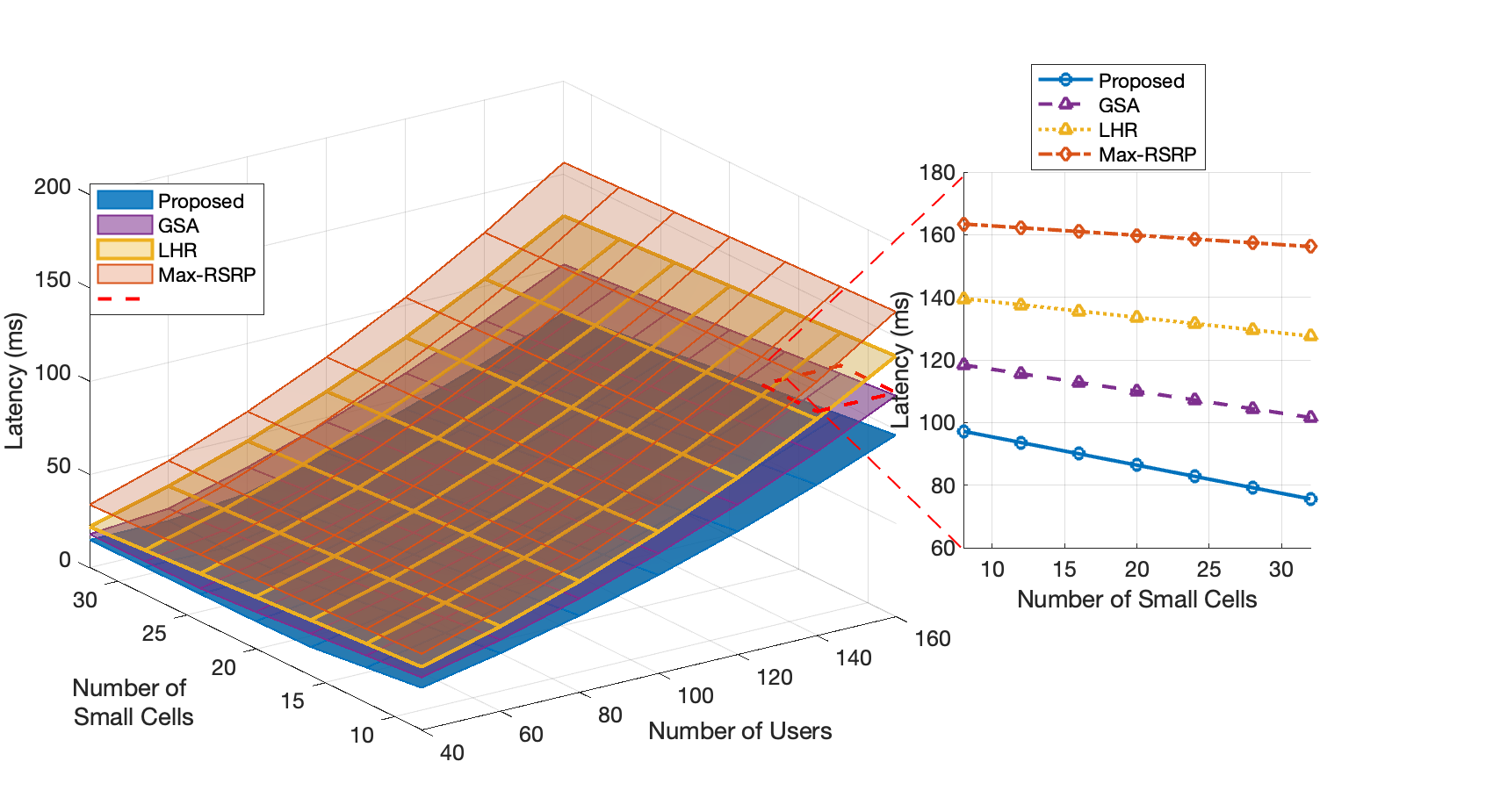}%
\label{fig6c}%
} %
\subfloat[]{%
\includegraphics[width=0.31\textwidth]{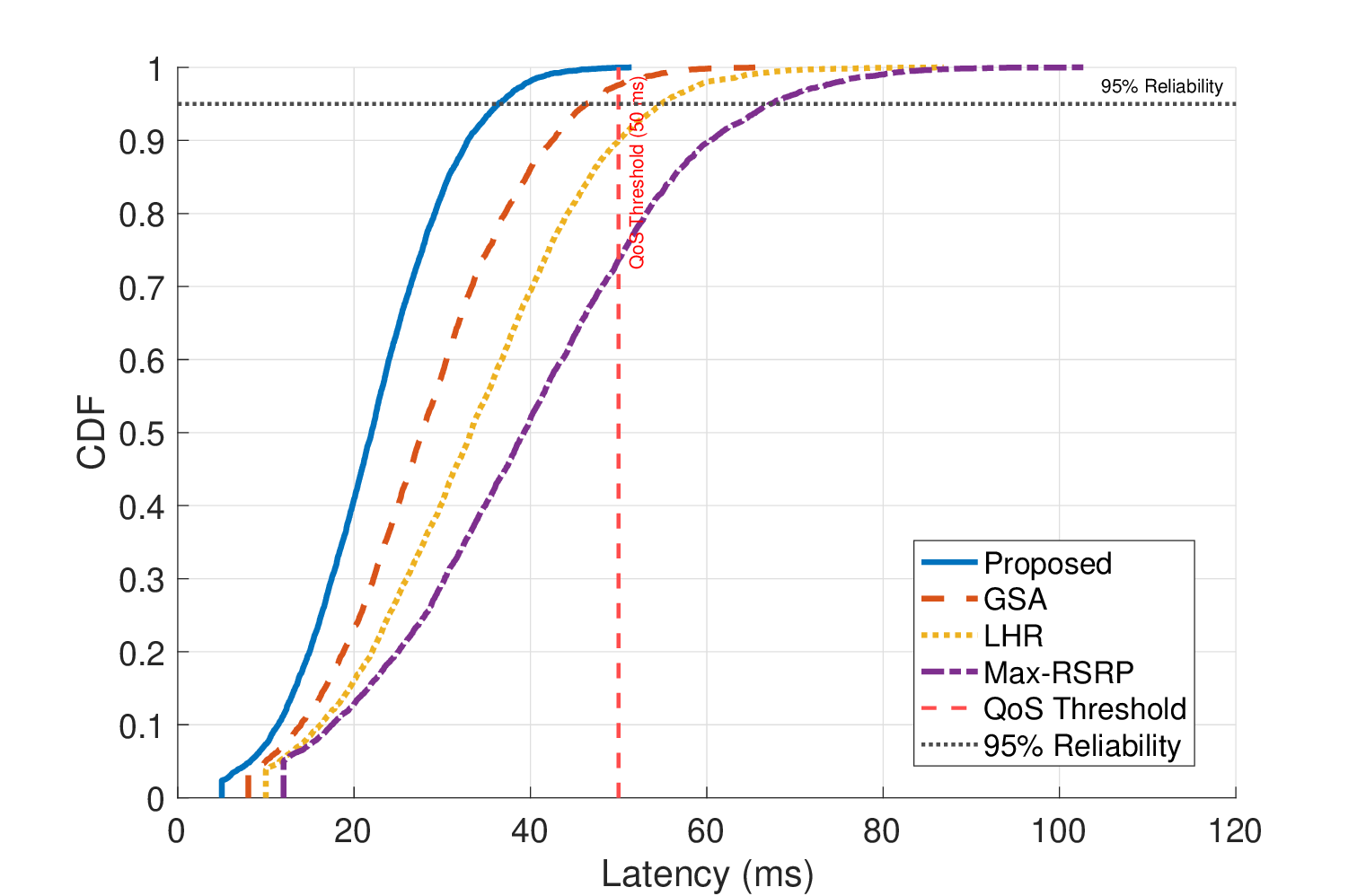}%

\label{f4}
} 
\caption{Latency performance under according to different parameters.}
\label{fig:latency_main}
\end{figure*}

\subsection{Performance Metrics}
To evaluate the effectiveness of the proposed framework, two key quality-of-service metrics are considered: average end-to-end latency and packet loss ratio. These metrics capture both the timeliness and reliability of packet delivery and are widely used in evaluating wireless access networks. All reported metrics are averaged over users, decision intervals, and independent simulation runs.

\begin{itemize}
\item \textit{Average End-to-End Latency:}
This metric represents the mean delay experienced by packets from the time they are generated at the source until they are successfully received at the destination. It includes queueing delay, transmission delay, propagation delay, and scheduling delay at the serving base station. The average latency is computed as $\bar{D} =
\frac{1}{MT}
\sum_{t=1}^{T}
\sum_{j=1}^{M}
D_{i^*(j,t),j}(t)$, where $M$ denotes the number of users, $T$ is the total number of decision intervals, and $D_{i^*(j,t),j}(t)$ represents the experienced end-to-end delay of user $j$ when associated with the selected small cell $i^*(j,t)$ at time interval $t$.
\item \textit{Packet Loss Ratio:}
This metric quantifies packet delivery reliability by measuring the fraction of packets that fail to reach the destination. Packet losses may occur due to buffer overflow, radio link failures, or scheduling limitations. The packet loss ratio is calculated as $\bar{P} =
\frac{
\sum_{t=1}^{T}
\sum_{j=1}^{M}
L_j(t)
}{
\sum_{t=1}^{T}
\sum_{j=1}^{M}
S_j(t)
}$, where $L_j(t)$ denotes the number of packets lost for user $j$ during interval $t$, and $S_j(t)$ represents the total number of packets generated for that user during the same interval.

\end{itemize}

\subsection{Simulation Results}

\subsubsection{Latency Evaluation}
Fig.~\ref{f1} illustrates the impact of user mobility on average end-to-end latency for all association schemes. Latency increases with mobility due to frequent handovers, channel variations, and transient queue imbalance. The proposed optimization–learning framework consistently achieves the lowest latency, with the performance gap widening at higher speeds. Unlike static or signal-strength-based approaches, it incorporates queue state, service degradation, and energy indicators to maintain balanced cell utilization under rapid topology changes. Among the baselines, GSA outperforms LHR and Max-RSRP at moderate speeds by considering multiple service indicators, but under high mobility it cannot anticipate post-handover congestion. LHR adapts more slowly to traffic shifts, while Max-RSRP exhibits the largest latency increase due to repeated associations with congested cells. Fig.~\ref{f3} shows average latency versus packet arrival rate under increasing traffic load. All schemes exhibit nonlinear delay growth due to queue buildup and resource contention. GSA performs well at light loads but experiences sharp latency spikes at high arrival rates, as short-term, greedy decisions fail to prevent congestion in heavily demanded cells. LHR partially considers load but lacks a global objective, leading to inefficient resource use under heavy traffic. Max-RSRP degrades the most, as signal-based association concentrates users on a subset of cells. In contrast, the proposed method maintains the lowest latency and increases more gradually, indicating that the optimization-guided learning mechanism distributes traffic more effectively while accounting for delay and congestion.

\begin{figure*}[h]
\centering
\subfloat[]{%
\includegraphics[width=0.31\textwidth]{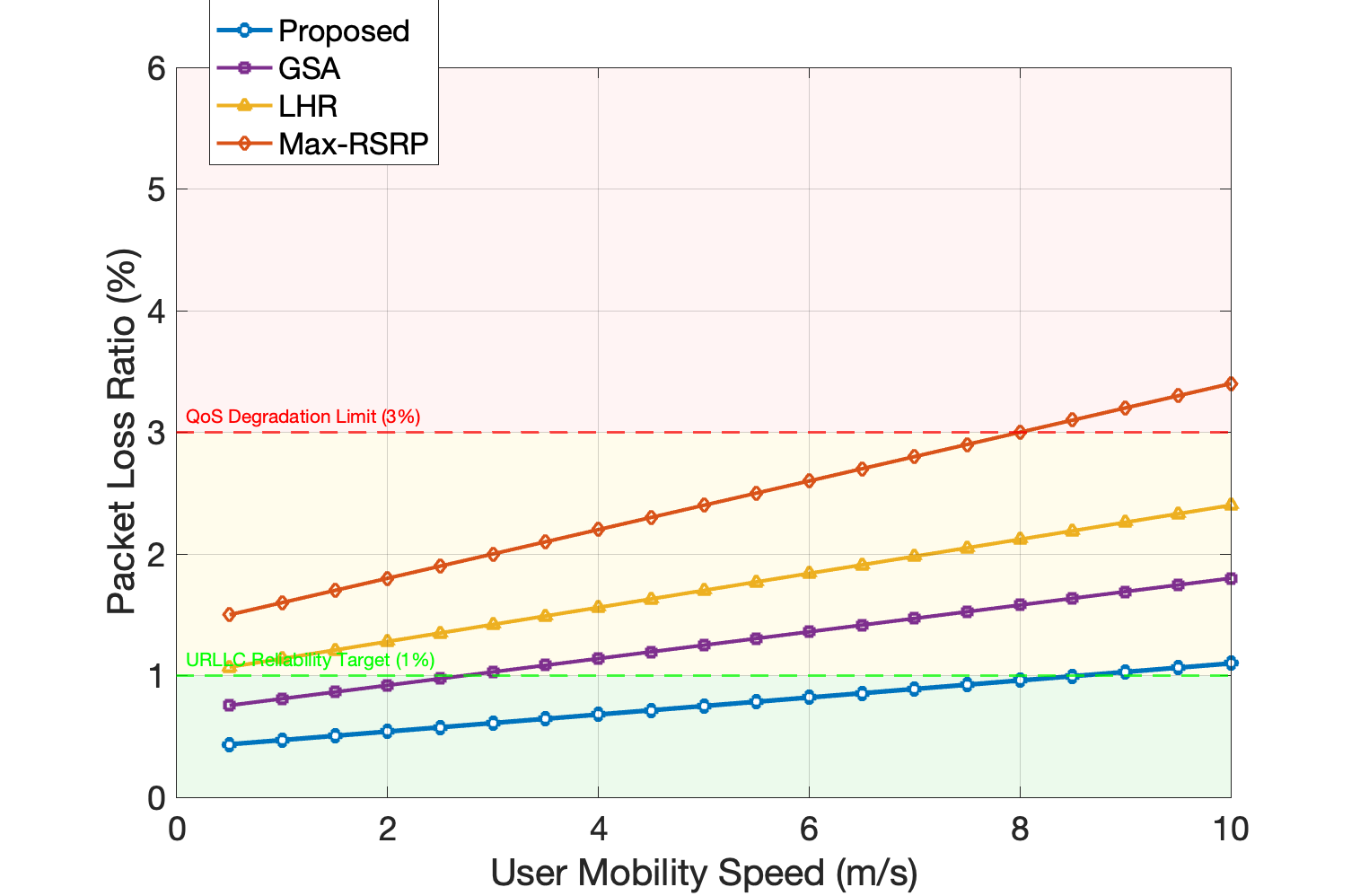}%
\label{f_plr_mobility}%
} %
\subfloat[]{%
\includegraphics[width=0.31\textwidth]{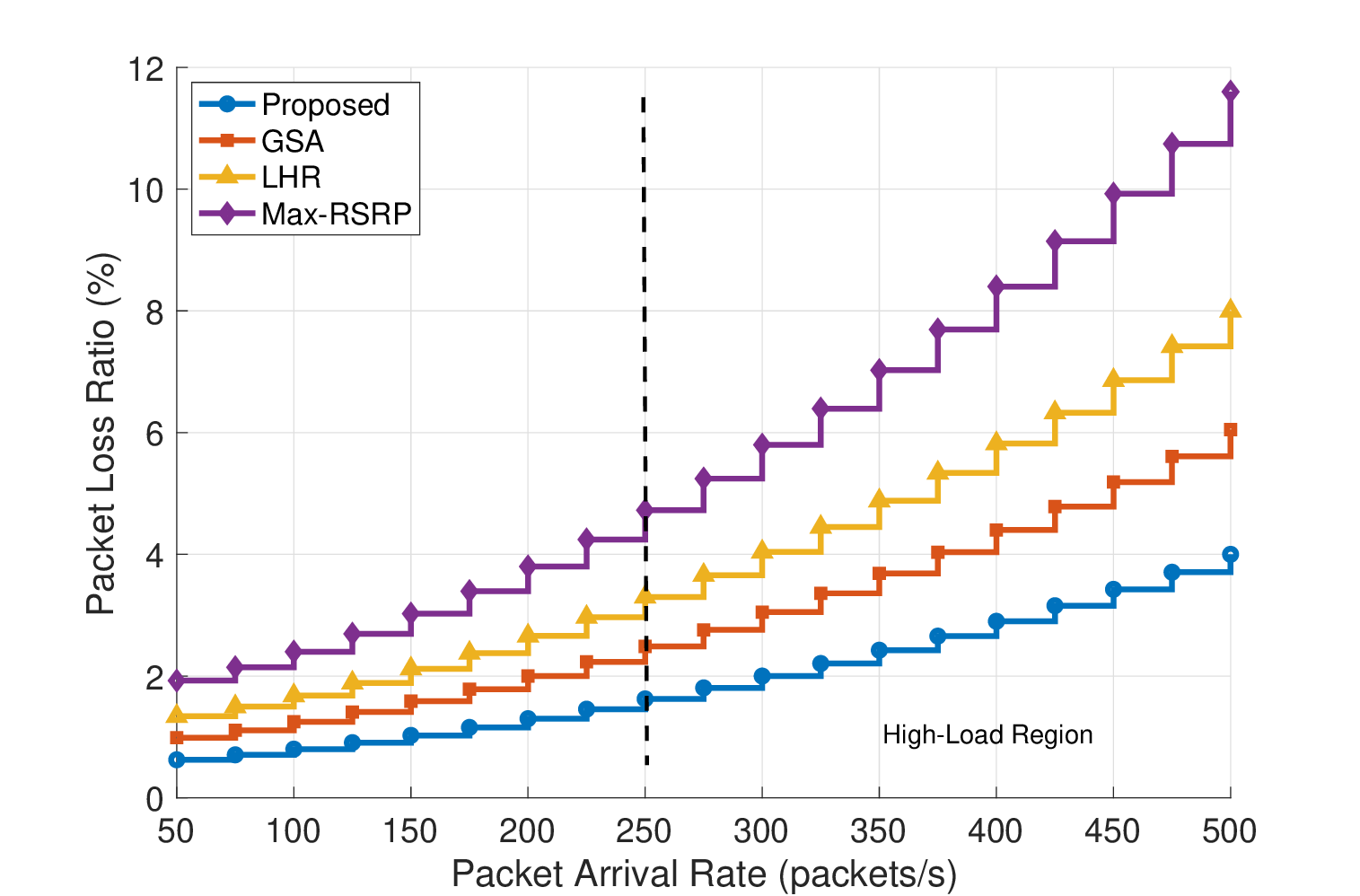}%
\label{f_plr_critical}%
} %
\subfloat[]{%
\includegraphics[width=0.31\textwidth]{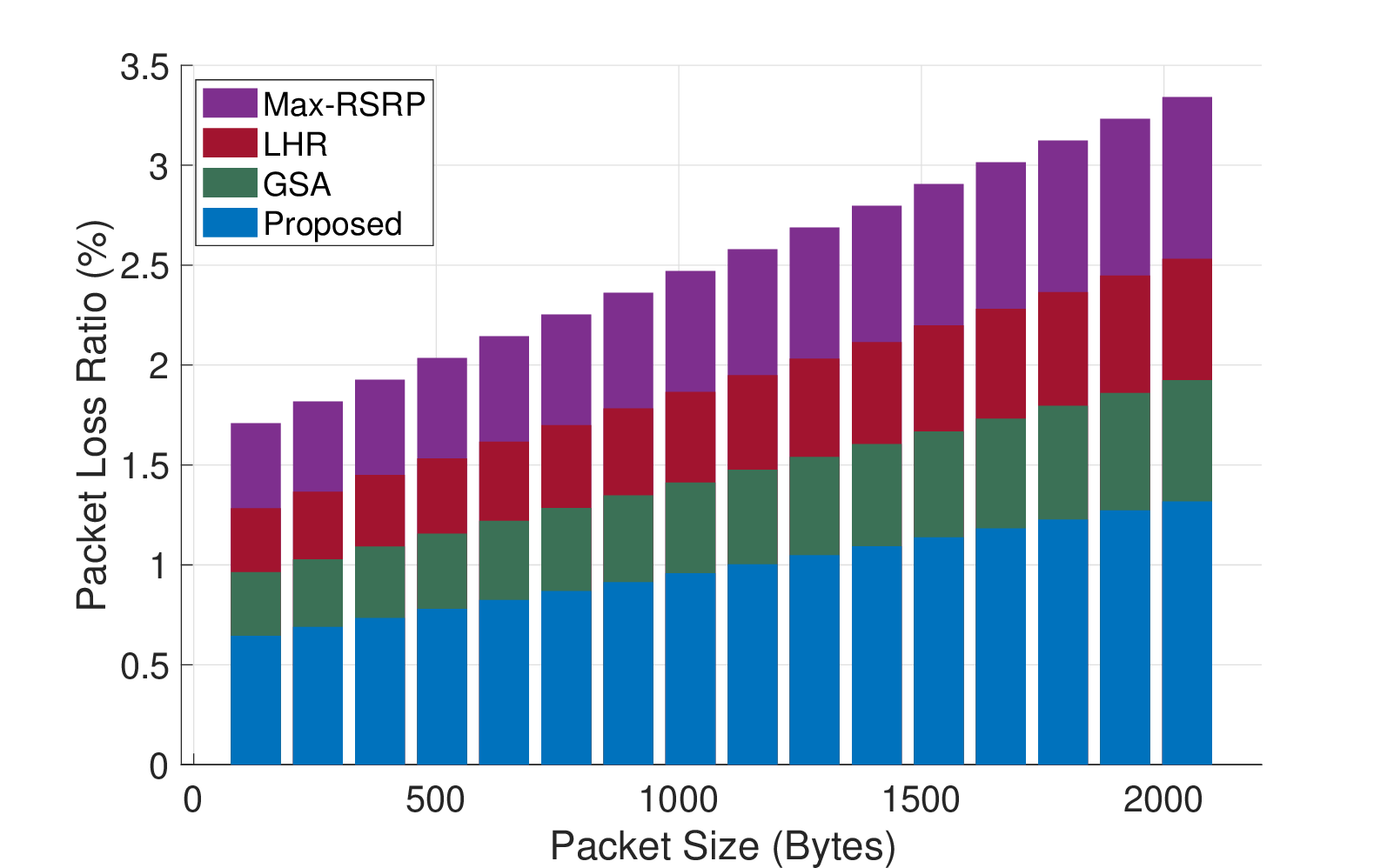}%
\label{f23}%
} \\
\subfloat[]{%
\includegraphics[width=0.46\textwidth]{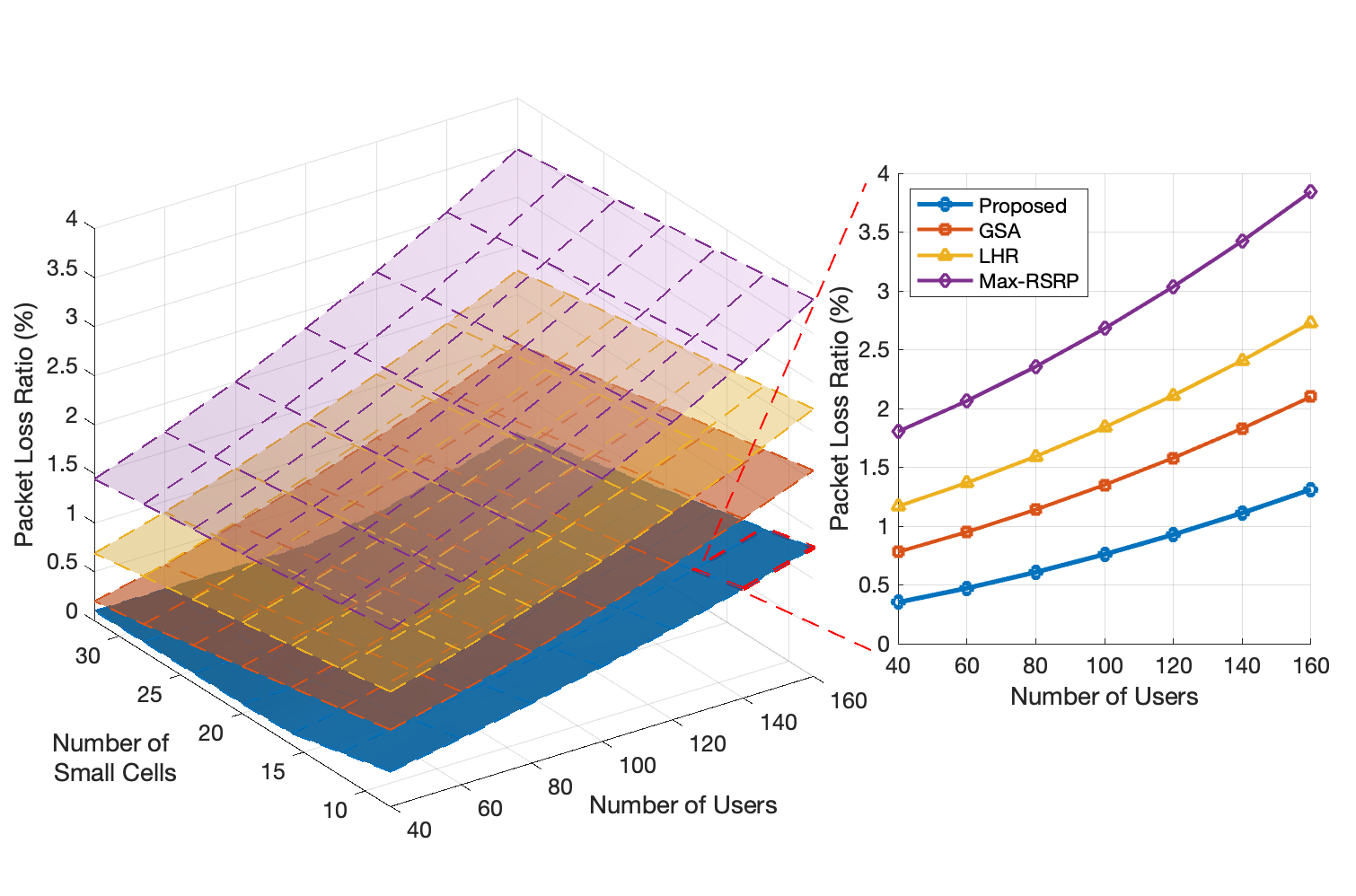}%
\label{f2}%
} %
\subfloat[]{%
\includegraphics[width=0.3\textwidth]{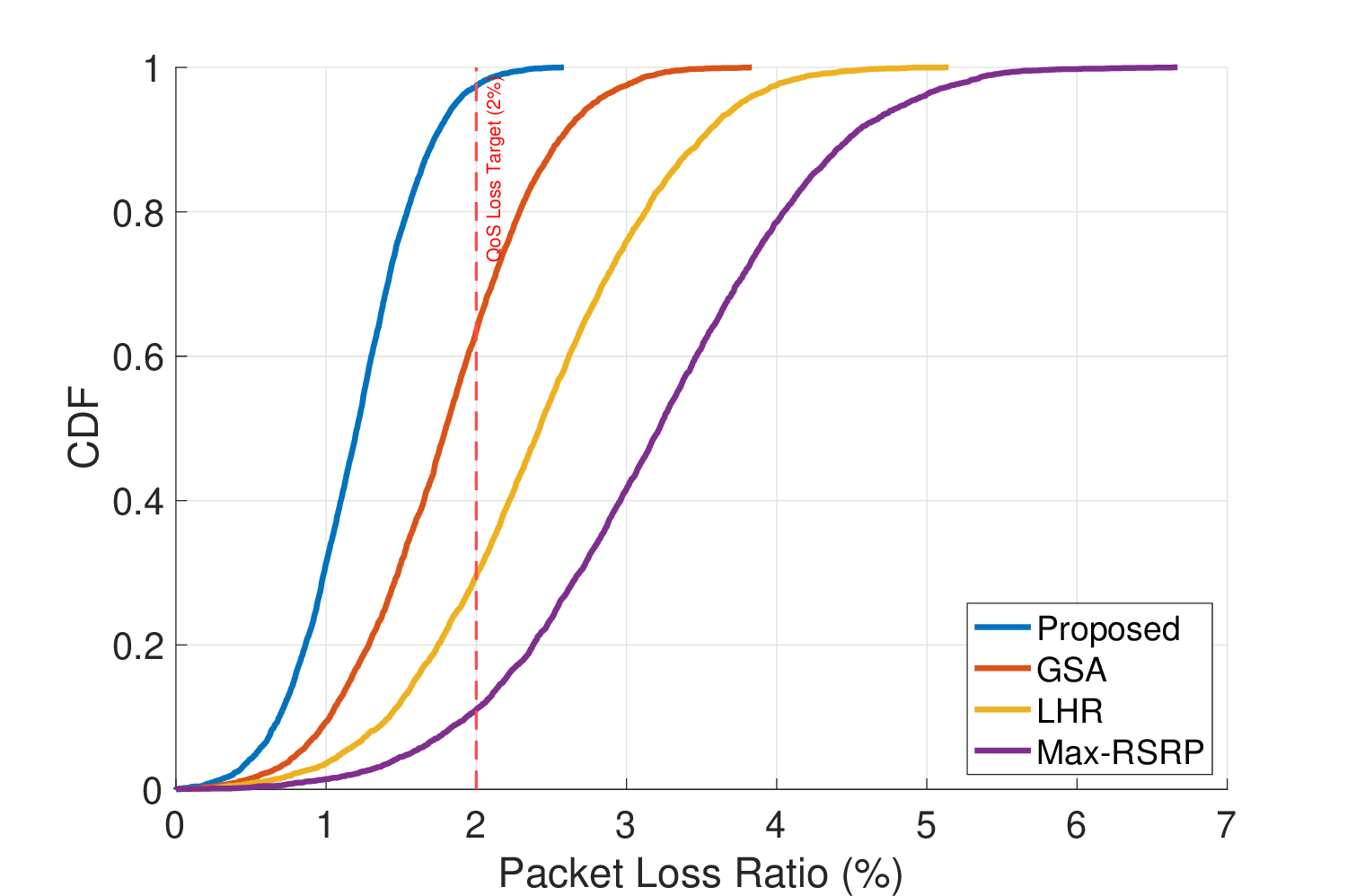}%
\label{f24}%
} \\
\caption{Analysis of packet loss according to different parameters.}
\label{r2}
\end{figure*}

A three-dimensional latency analysis with respect to network density is presented in Fig.~\ref{fig6c}, where density reflects different combinations of user and small-cell counts. Latency increases with the number of users due to higher offered load, while additional small cells generally reduce delay through spatial reuse and improved load balancing. 
\begin{table}[h]
\centering
\caption{Latency improvement of the proposed method.}
\label{tab:latency_improvement}
\scriptsize
\renewcommand{\arraystretch}{1}
\begin{tabular}{lccc}
\toprule
\textbf{Scenario} & \textbf{vs. GSA} & \textbf{vs. LHR} & \textbf{vs. Max-RSRP} \\
\midrule
High Mobility (3 m/s) & 18.7\% & 26.5\% & 41.9\% \\
High Traffic Load (400 kbps) & 22.9\% & 30.8\% & 46.0\% \\
Dense Network (160 users) & 22.0\% & 31.9\% & 44.2\% \\
Large Packet Size (1408 B) & 32.0\% & 20.9\% & 47.7\% \\
95\% Reliability Latency & 17.2\% & 25.0\% & 43.5\% \\
\bottomrule
\end{tabular}
\end{table} The proposed framework achieves the lowest latency across all configurations and scales smoothly with both user growth and network densification. GSA benefits moderately from densification but still suffers from localized congestion, whereas LHR gains less due to its limited load metric. Max-RSRP remains highly sensitive to user density because radio-centric decisions fail to utilize newly deployed cells efficiently. Fig.~\ref{f5} further shows latency variation with packet size. Larger packets increase transmission and service time, leading to higher delays for all schemes. Nevertheless, the proposed framework degrades more gracefully and maintains latency below the 50 ms QoS target across a wider packet-size range. Finally, Fig.~\ref{f4} presents the cumulative distribution function (CDF) of user latency. The 50~ms QoS threshold and the 95\% reliability marker indicate the probability that user delay remains within the target bound. The proposed method produces the leftmost and steepest curve, reflecting both lower latency and reduced variability. More than 95\% of users remain below the QoS threshold, while GSA, LHR, and especially Max-RSRP exhibit wider delay distributions. Table~\ref{tab:latency_improvement} summarizes the latency reduction achieved by the proposed framework under representative scenarios. The improvements remain consistent across scenarios and become particularly evident under high mobility and heavy traffic. The method also scales better with user density and larger packet sizes. Additionally, the 95th-percentile latency is significantly reduced, confirming that the gains extend beyond average delay to overall service reliability.

\subsubsection{Packet Loss Evaluation}

Fig.~\ref{f_plr_mobility} presents the variation in packet loss ratio (PLR) with user mobility speed, along with service-oriented reliability regions. The shaded bands distinguish ultra-reliable ($\leq 1\%$), moderate (1–3\%), and degraded ($\geq 3\%$) regimes, enabling clearer interpretation of reliability trends. As mobility increases, PLR rises across all schemes due to more frequent handovers, rapid channel changes, and temporary queue imbalances. The proposed optimization–learning framework shows the slowest PLR growth and remains within the ultra-reliable region over a wider mobility range than the baselines. This suggests that mobility-aware load redistribution and queue-sensitive decisions effectively reduce transient congestion. GSA initially maintains moderate reliability but degrades at higher speeds because short-term decisions cannot stabilize post-handover load. LHR adapts more slowly to traffic shifts, while Max-RSRP enters the degraded region first, highlighting the limitations of signal-strength-based association.

Fig.~\ref{f_plr_critical} depicts PLR versus packet arrival rate and identifies the traffic level at which each scheme violates the 2\% QoS loss constraint, highlighting congestion tolerance rather than simple monotonic degradation. The proposed framework sustains acceptable PLR even at high arrival rates, demonstrating strong resilience to traffic surges. GSA violates the threshold earlier due to overload at frequently selected cells, LHR fails at lower traffic because of its simplified load metric, and Max-RSRP exhibits the earliest violation.

\begin{table}[h]
\centering
\caption{Packet loss reduction of the proposed method.}
\label{tab:plr_improvement}
\scriptsize % \scriptsize yerine daha okunabilir olan \small önerilir
\renewcommand{\arraystretch}{1.1}
\begin{tabular}{lccc}
\toprule
\textbf{Scenario} & \textbf{vs. GSA} & \textbf{vs. LHR} & \textbf{vs. Max-RSRP} \\
\midrule
High Mobility (3 m/s)        & 21.4\% & 33.8\% & 48.6\% \\
High Traffic Load (400 kbps) & 24.7\% & 36.2\% & 52.9\% \\
Dense Network (160 users)    & 19.8\% & 31.5\% & 45.1\% \\
Large Packet Size (1408 B)   & 23.6\% & 34.4\% & 50.7\% \\
90\% Reliability PLR         & 18.9\% & 29.6\% & 44.8\% \\
\bottomrule
\end{tabular}
\end{table}

The joint effect of user density and small-cell deployment is illustrated by the three-dimensional PLR surface in Fig.~\ref{f2}. PLR increases with user density due to contention and queue buildup, while densification reduces loss through improved spatial reuse and load balancing. The proposed framework achieves the lowest PLR across all configurations and scales smoothly with both user growth and network densification. GSA benefits from additional cells but remains sensitive to user growth; LHR shows limited improvement; and Max-RSRP exhibits the steepest PLR escalation and the weakest densification gain.

Fig.~\ref{f23} evaluates the impact of packet size. Larger packets increase channel occupancy and buffer pressure, thereby increasing PLR across all schemes. The proposed method degrades most gradually, indicating effective queue-aware congestion mitigation. GSA degrades with larger packets, while LHR and especially Max-RSRP show higher sensitivity. The user-level reliability distribution in Fig.~\ref{f24} further confirms this trend: the proposed scheme yields the steepest and leftmost CDF, with over 90\% of users remaining below the 2\% loss threshold, whereas baseline methods exhibit broader distributions and heavier loss tails.

Table~\ref{tab:plr_improvement} quantifies PLR reductions under representative stress scenarios. The gains remain consistent across variations in mobility, load, density, and packet size, and are most pronounced under heavy traffic and large packets where congestion dominates. Improvements in high-percentile PLR further demonstrate enhanced reliability stability across users.

\subsection{System Overhead, Complexity Evaluation, and Ablation Study}
\begin{figure}[h]
\centering
\includegraphics[width=\linewidth]{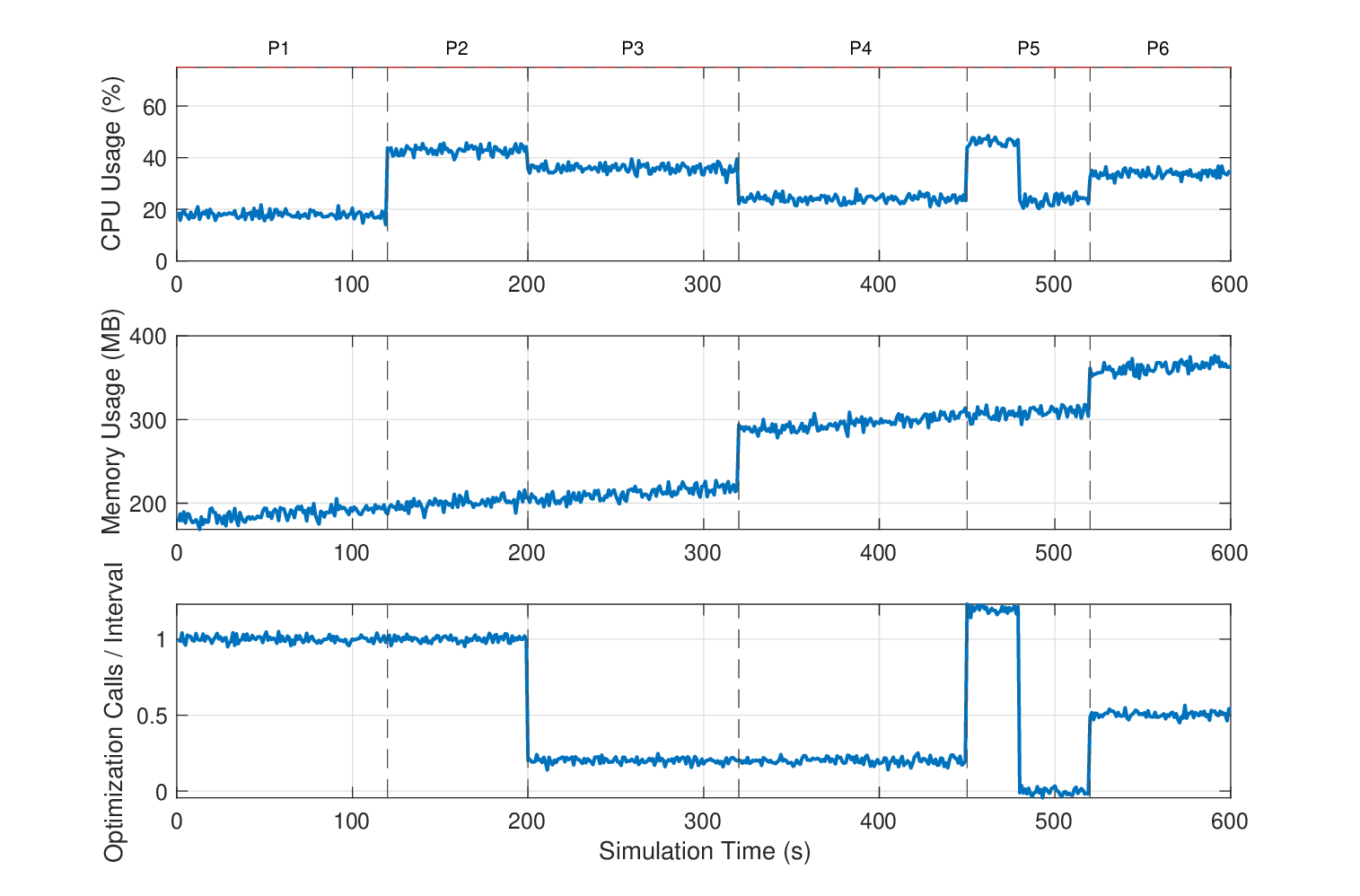}
\caption{System overhead during simulation time}
%\vspace{-2em}
\label{fig:overhead}
\end{figure}

Beyond network-level performance metrics, Fig.~\ref{fig:overhead} illustrates the temporal evolution of computational overhead in the proposed framework. The operational timeline is divided into several phases that reflect how the hybrid optimization–learning architecture behaves during deployment. The process begins with a lightweight monitoring phase (P1), during which the system collects queueing, delay, and service indicators from small cells. Since this stage mainly involves measurement and aggregation operations, CPU utilization remains relatively low. However, memory usage gradually increases as the system accumulates state observations and network statistics that will later be used by the learning module. Following this stage, the framework enters the optimization-driven control phase (P2), where the Lagrangian optimization engine is executed repeatedly to generate high-quality user–cell association decisions. This phase produces the highest CPU utilization because the algorithm iteratively updates dual variables and evaluates association costs across all potential user–cell pairs. Although computationally intensive, this phase is temporary and primarily serves to generate reliable supervisory labels for the subsequent learning stage. Next, the system transitions to the learning phase (P3), during which the LVQ model is trained using the optimization outputs. Compared with iterative optimization, this stage results in more stable yet moderately elevated CPU usage due to prototype adaptation, feature processing, and training updates. Memory consumption also increases as labeled samples and prototype vectors are stored in the knowledge base.

After convergence, the framework moves to inference-dominant operation (P4). In this phase, most optimization calls are replaced by lightweight LVQ-based classification. As a result, CPU utilization drops significantly while maintaining association decisions that closely approximate the optimization solution. This transition highlights the core benefit of the hybrid design: computationally expensive optimization is replaced by efficient learning-based inference during steady-state operation. To maintain robustness under evolving traffic patterns, periodic re-optimization bursts are triggered (P5). These appear as short CPU spikes in Fig.~\ref{fig:overhead}, but they occur infrequently and therefore do not significantly impact the long-term computational profile. Finally, under sustained heavy traffic conditions (P6), monitoring and inference costs increase slightly due to higher data-plane activity and more frequent state updates. Nevertheless, overall resource consumption remains well within practical limits, indicating that the proposed architecture is suitable for real-time operation in dense small-cell networks.

\begin{table}[h]
\centering
\caption{Computational complexity of components.}
\label{tab:complexity}
\scriptsize
\renewcommand{\arraystretch}{1}
\begin{tabular}{lll}
\toprule
\textbf{Module} & \textbf{Time Complexity} & \textbf{Execution Frequency} \\
\midrule
Monitoring & $\mathcal{O}(N_s M)$ & Every decision interval \\
Optimization & $\mathcal{O}(I \cdot N_s M)$ & Periodic (re-opt phase) \\
Feature Processing & $\mathcal{O}(N_s M)$ & During training/inference \\
LVQ Training & $\mathcal{O}(K d)$ & Initial + periodic \\
LVQ Inference & $\mathcal{O}(K d)$ per candidate cell & Most decision intervals \\
Knowledge Storage & $\mathcal{O}(N_s M)$ & Every decision interval \\
\bottomrule
\end{tabular}
\end{table}

Table~\ref{tab:complexity} summarizes the computational complexity of the main framework phases. Monitoring and feature construction scale linearly with the number of user–cell pairs, yielding $\mathcal{O}(N_s M)$ complexity, since service indicators must be evaluated for each candidate association. Here, $N_s$ and $M$ denote the numbers of small cells and users, respectively. These operations are lightweight and executed continuously as part of network telemetry. The Lagrangian optimization module incurs the highest cost, $\mathcal{O}(I \cdot N_s M)$, where $I$ is the number of convergence iterations. However, this module operates only during initialization and periodic re-optimization, so its cost is amortized over time rather than incurred at every decision interval. In contrast, the LVQ learning component is significantly lighter. Both training and inference require $\mathcal{O}(K d)$ computations, where $K$ is the number of prototypes and $d$ the feature dimension. During online operation, inference is performed for each candidate cell, leading to $\mathcal{O}(N_s K d)$ complexity per interval. Since $K$ and $d$ remain small and independent of network scale, inference overhead is substantially lower than repeated global optimization. The knowledge storage module introduces additional linear overhead, $\mathcal{O}(N_s M)$, for recording state–action pairs and service indicators. As this primarily involves memory operations, it does not affect real-time latency. Therefore, the proposed framework shifts computational burden from continuous large-scale optimization to lightweight prototype-based inference. Global optimization is invoked intermittently to refresh the knowledge base, while most decisions rely on efficient learning-based classification, enabling near-optimal performance with real-time scalability in dense 6G small-cell deployments.

\begin{figure}[h]
\centering
\includegraphics[width=\linewidth]{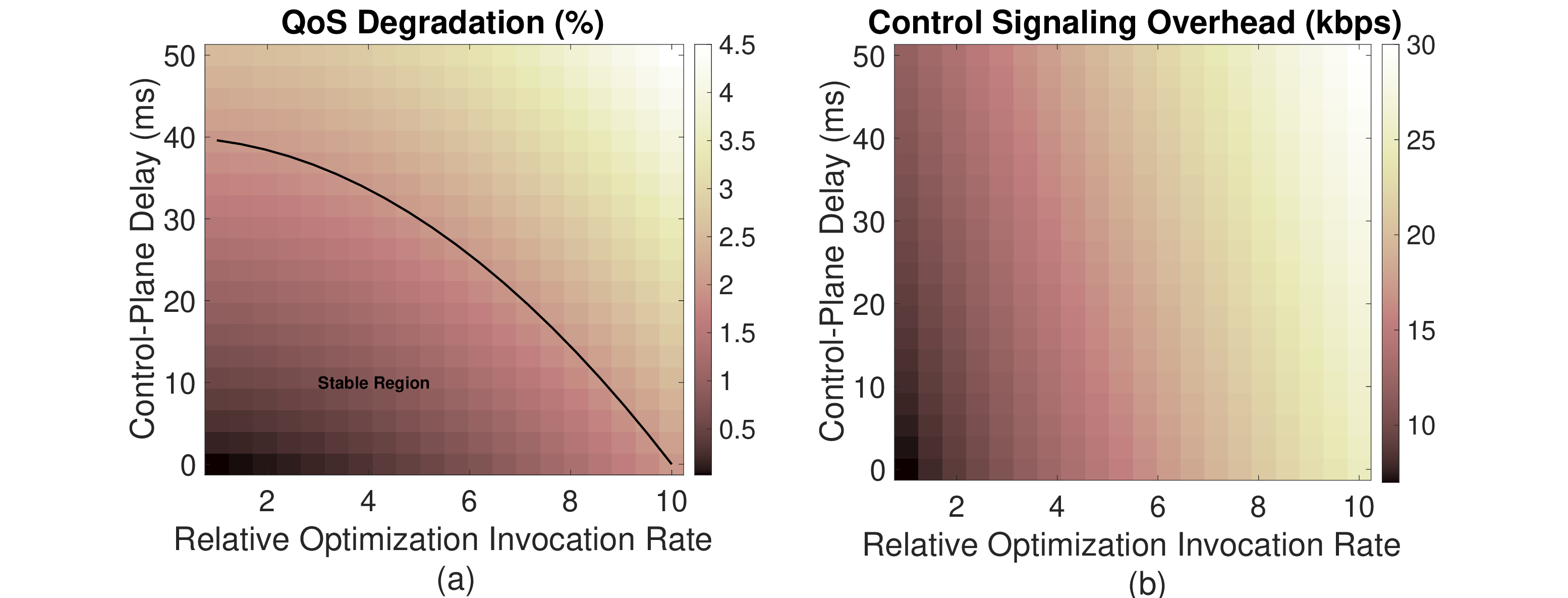}
\caption{Control-Loop Overhead and Adaptation Cost Analysis}
%\vspace{-2em}
\label{fig:control_tradeoff}
\end{figure}
Beyond computational complexity, the practical feasibility of the proposed KDN framework depends on two control-loop factors: (i) the invocation frequency of the global optimization module, and (ii) the latency between the knowledge and data planes. Excessively frequent optimization increases computational and signaling overhead, whereas infrequent invocation may delay adaptation to traffic fluctuations and mobility dynamics. To analyze this trade-off, different optimization intervals were emulated under fixed traffic and mobility conditions. The resulting QoS degradation, defined as the relative increase in the composite latency–packet-loss metric compared with continuous optimization, is shown in Fig.~\ref{fig:control_tradeoff}a. Within moderate invocation intervals, QoS degradation remains below approximately 3\%, while substantially reducing the number of global optimization executions. This confirms that the hybrid optimization–learning design effectively amortizes computational cost while preserving service quality.

The second deployability factor is control-plane signaling delay, which introduces a gap between state observation and enforcement. Controlled delays were injected into the decision loop, and the results are presented in Fig.~\ref{fig:control_tradeoff}b. Even with 30–40 ms of delay, additional QoS degradation remains below approximately 2\%. This limited sensitivity stems from the relatively slower time scale of queue evolution and traffic redistribution in dense small-cell networks. Consequently, moderate control-plane delays do not significantly affect association stability, indicating that the proposed KDN framework can operate reliably under realistic latency conditions.

\begin{figure}[h]
\centering
\subfloat[]{%
\includegraphics[width=0.24\textwidth]{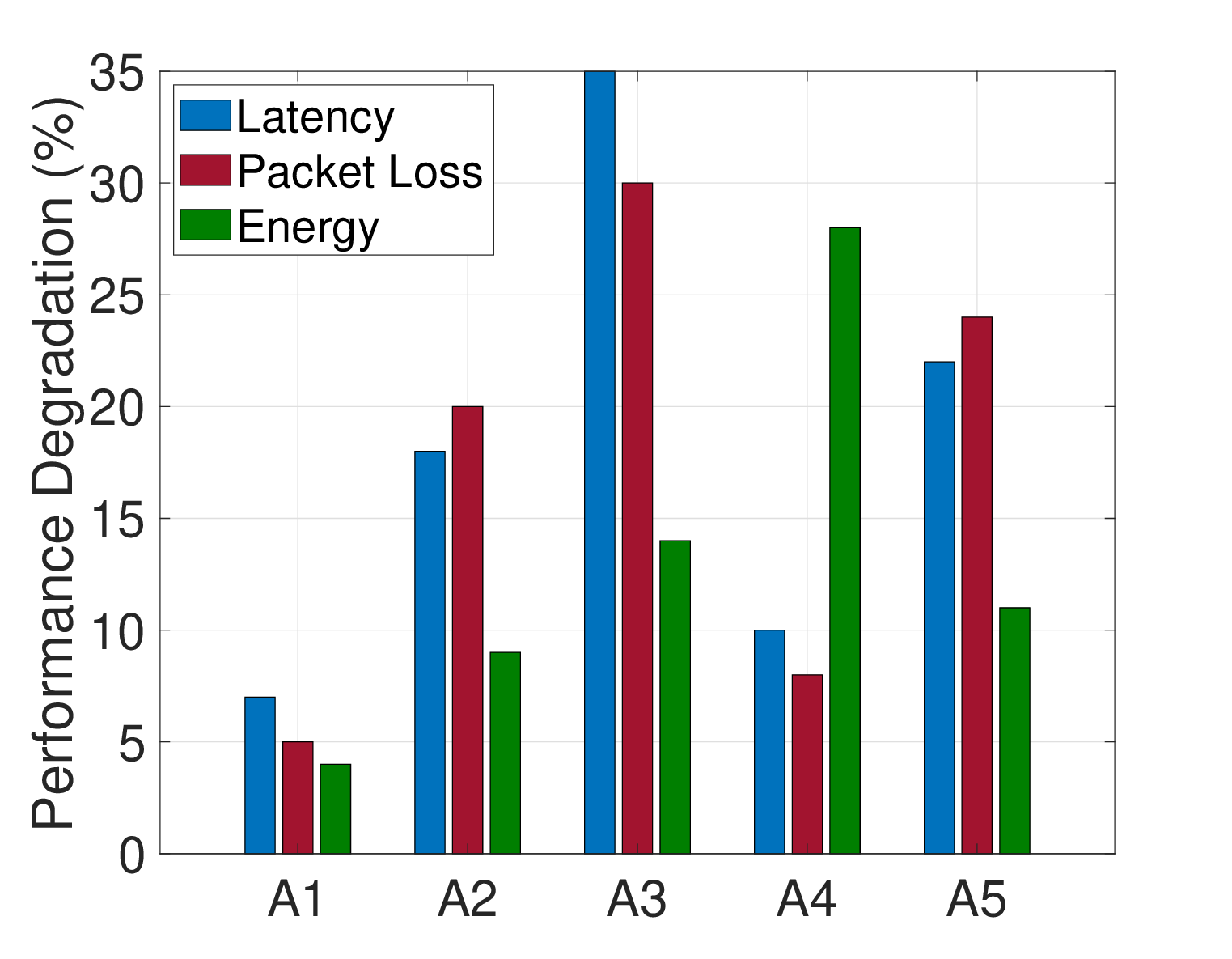}%
\label{fig:ablation}%
} %
\subfloat[]{%
\includegraphics[width=0.24\textwidth]{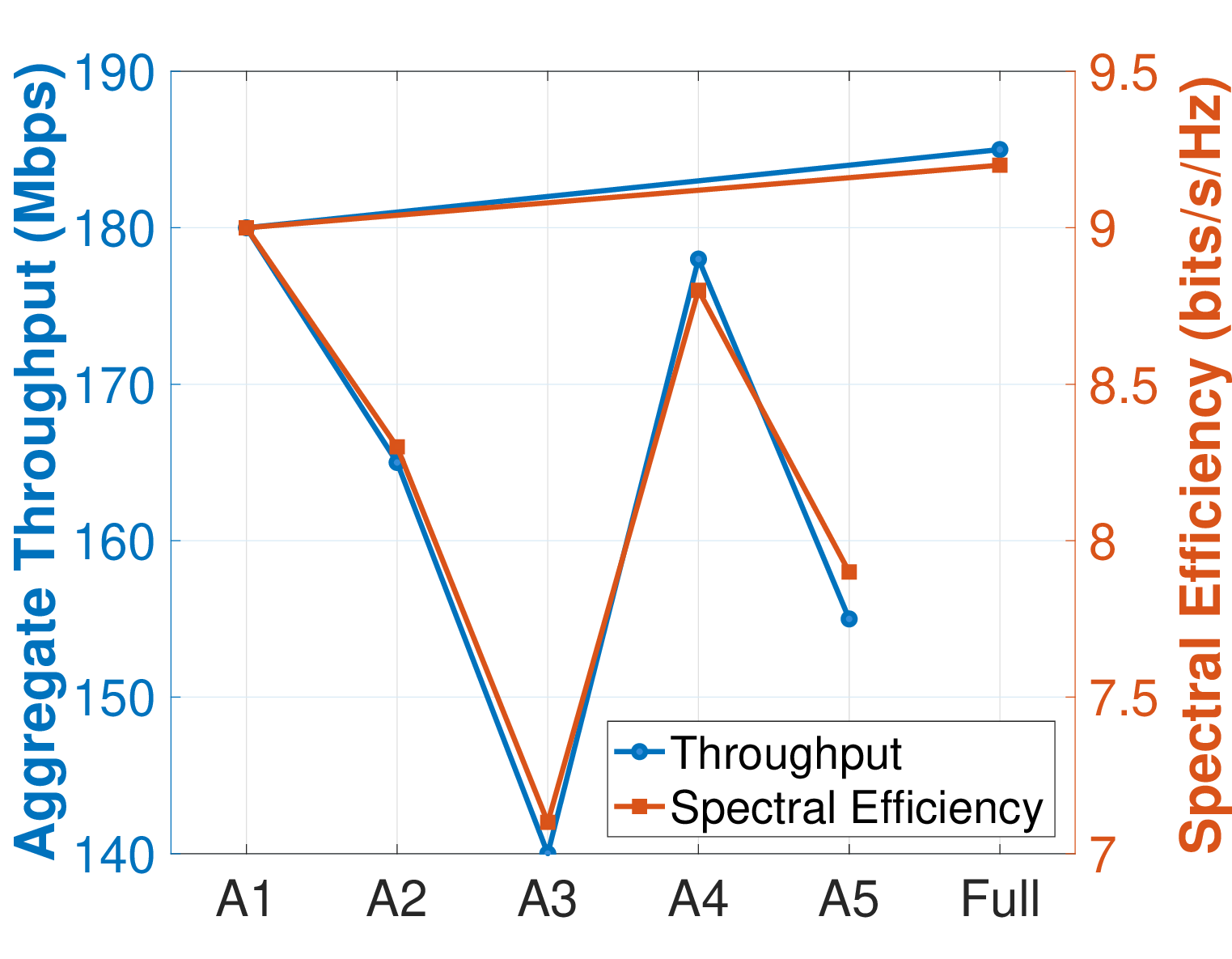}%
\label{fig:ablation_efficiency}%
} \
\caption{Ablation study evaluations.}
\label{r2}
\end{figure}

Moreover, Fig.~\ref{fig:ablation} presents an ablation study illustrating the contribution of each module, where each variant disables a single component while preserving the others. The most severe degradation occurs in Variant A3 (no queue-awareness), with latency and packet loss increasing by more than 30\%, confirming that congestion awareness is the dominant QoS factor in dense small-cell environments. Without queue-state information, the association process cannot react to load variations, leading to persistent imbalance and buffer accumulation. Variant A5 (no service degradation indicators) produces the second-largest deterioration, highlighting the importance of delay- and packet-related metrics. Removing optimization-generated supervisory labels (Variant A2) also degrades performance, as the learning module then operates without global guidance. In contrast, eliminating energy awareness (Variant A4) mainly increases total energy consumption with limited QoS impact, while disabling LVQ (Variant A1) causes only minor QoS loss but significantly increases computational complexity, confirming its role as an acceleration layer.

Fig.~\ref{fig:ablation_efficiency} complements the QoS analysis by evaluating aggregate throughput and spectral efficiency. The complete framework achieves the highest performance, indicating improvements in both QoS and global resource utilization. The largest efficiency loss again appears in Variant A3, where load imbalance reduces throughput by nearly 25\%. Variants A2 and A5 also reduce efficiency, whereas removing LVQ has minor throughput impact but higher computational cost, and excluding energy awareness mainly affects power distribution. Overall, the results confirm that queue-awareness and service-quality indicators are the most critical components for sustaining QoS and efficiency, while the learning module enables scalable implementation with minimal performance loss.

\begin{figure*}[h]
\centering
\hspace{-0.4in}
\subfloat[]{%
\includegraphics[width=0.51\textwidth]{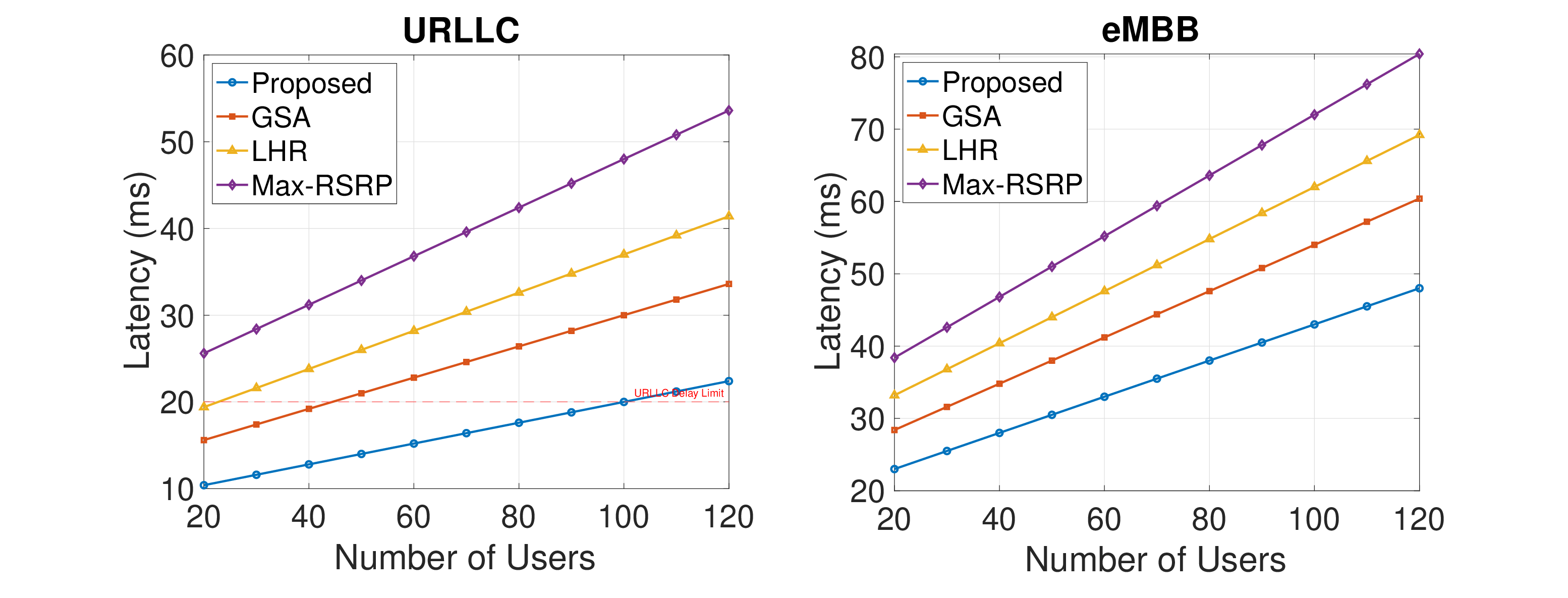}%
\label{fig:service_latency}%
} %
\hspace{-0.3in}
\subfloat[]{%
\includegraphics[width=0.51\textwidth]{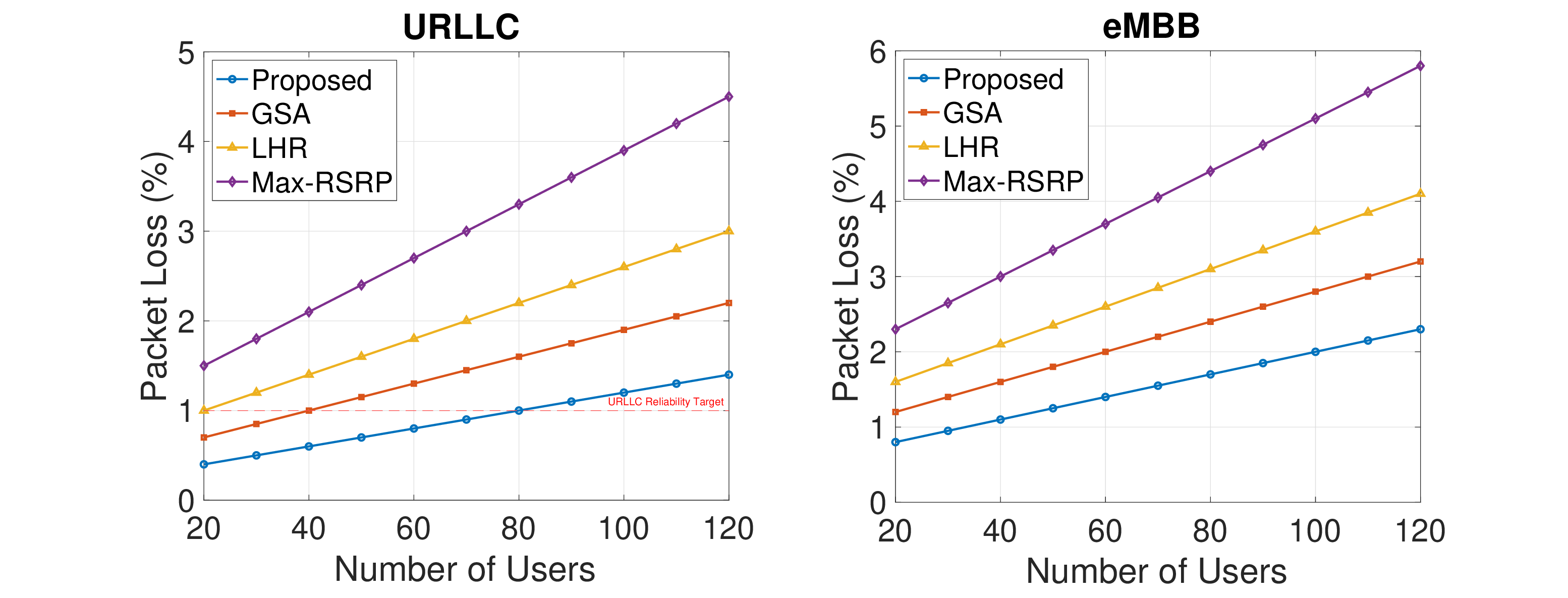}%
\label{fig:service_loss}%
} \\
\caption{Performance evaluation for heterogeneous service classes}
\label{comparison}
\end{figure*}

\begin{table*}[h]
\centering
\caption{Comparison of candidate classifiers for edge deployment.}
\label{tab:model_compare}
\scriptsize
\renewcommand{\arraystretch}{1}
\begin{tabular}{lllll}
\toprule
\textbf{Method} & \textbf{Training Complexity} & \textbf{Inference Complexity} & \textbf{Interpretability} & \textbf{Scalability w.r.t. $N$} \\
\midrule
LVQ & $\mathcal{O}(N K d)$ & $\mathcal{O}(K d)$ & High (prototype-based) & Independent at inference \\
KNN & $\mathcal{O}(1)$ & $\mathcal{O}(N d)$ & Medium & Linear growth \\
MLP (1 hidden layer) & $\mathcal{O}(N d h)$ & $\mathcal{O}(d h)$ & Low & Independent at inference \\
Logistic Regression & $\mathcal{O}(N d)$ & $\mathcal{O}(d)$ & High & Independent at inference \\
Decision Tree & $\mathcal{O}(N d \log N)$ & $\mathcal{O}(\log N)$ & Medium & Depth-dependent \\
\bottomrule
\end{tabular}
\end{table*}

\vspace{-0.19in}
\subsection{Justification of the Selected Learning Model}

Although the optimization framework yields analytically grounded association decisions, a surrogate classifier is required for low-latency edge inference. The selected model must satisfy four constraints: (i) bounded inference complexity, (ii) latency independent of dataset size, (iii) moderate memory footprint, and (iv) sufficient nonlinear capacity to approximate the optimization policy. LVQ is adopted as an optimizer-guided supervised classifier, where labels are generated by the Lagrangian optimization module. This preserves the structural properties of the original objective while substantially reducing runtime overhead.

\begin{figure}[h]
\centering
\includegraphics[width=\linewidth]{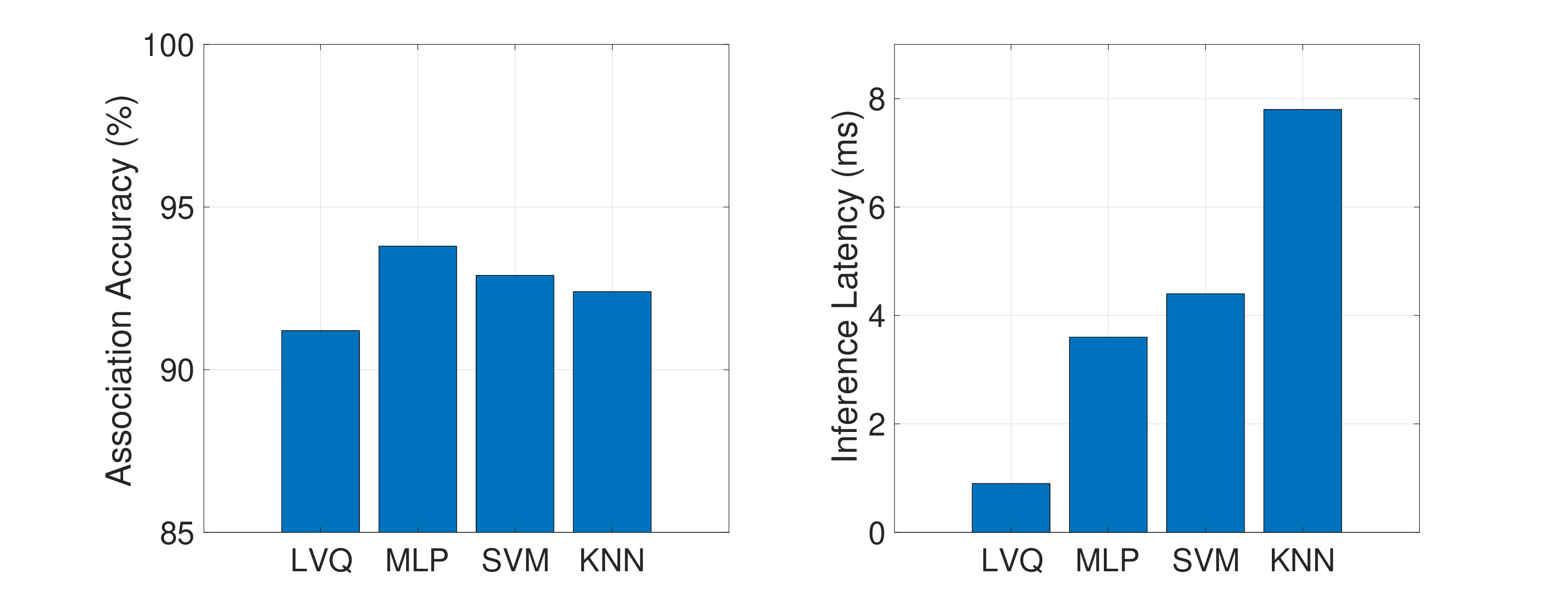}
\caption{Comparison of candidate machine learning models.}
\label{machine}
\end{figure}

Let $N$ denote the number of training samples, $K$ the number of prototypes, and $d$ the feature dimension. LVQ training scales as $\mathcal{O}(N K d)$, while inference requires only $\mathcal{O}(K d)$ prototype-distance evaluations. Since $K \ll N$, inference complexity remains bounded and independent of dataset size after training, ensuring deterministic latency suitable for CPU-based edge controllers without hardware acceleration. Alternative lightweight classifiers present different trade-offs. KNN incurs $\mathcal{O}(N d)$ inference complexity due to full-dataset distance computation, leading to dataset-dependent latency growth. Shallow MLPs offer higher representational capacity but increase inference latency and memory footprint due to hidden-layer matrix multiplications. Logistic regression has low complexity but limited ability to model nonlinear congestion–delay–energy interactions. Decision trees may exhibit depth-dependent latency and sensitivity to distributional shifts. Table~\ref{tab:model_compare} summarizes the theoretical comparison, showing that LVQ uniquely combines bounded inference complexity, prototype-level interpretability, and stable scaling with dataset growth. Fig.~\ref{machine} illustrates the empirical latency–accuracy trade-off across candidate models. While shallow neural networks achieve marginally higher classification accuracy, they incur significantly higher inference latency. KNN maintains competitive accuracy but exhibits the largest latency due to dataset-dependent distance computations. In contrast, LVQ operates in the low-latency region of the trade-off curve while maintaining accuracy close to neural baselines, confirming that it provides the most favorable balance between decision quality and real-time applicability in dense 6G small-cell environments. These theoretical and empirical results justify LVQ as an efficient and scalable surrogate model for edge-based small-cell selection.

\vspace{-0.15in}
\subsection{Performance under Heterogeneous Service Classes}
To evaluate service awareness, we examine the framework under heterogeneous traffic classes representing URLLC and eMBB services, which impose distinct QoS requirements. URLLC prioritizes ultra-low latency and high reliability, whereas eMBB primarily targets high throughput with more relaxed delay constraints. In the simulations, URLLC traffic uses small packets (64 bytes), strict delay sensitivity, and high scheduling priority, while eMBB flows employ larger packets (1024 bytes) and higher data rates. Both traffic types coexist and compete for shared radio and queue resources. Fig.~\ref{fig:service_latency} and Fig.~\ref{fig:service_loss} present the resulting latency and packet loss. The proposed optimization--learning framework provides the largest improvement for URLLC traffic. Because URLLC is highly sensitive to queue buildup and transient congestion, incorporating queue-state and service indicators enables the system to avoid overloaded cells more effectively, keeping latency and loss relatively stable as load increases. In contrast, baseline schemes exhibit stronger degradation because their decisions rely primarily on radio metrics or simplified load information. For eMBB traffic, performance differences are smaller but still visible. As eMBB tolerates higher delay, the gains mainly appear as reduced packet loss and more balanced traffic distribution across cells. These results indicate that the proposed framework adapts association decisions according to service characteristics while maintaining efficient resource utilization in dense 6G environments.

\vspace{-0.1in}
\subsection{Optimization--Learning Gap and Convergence Behavior}
While integrating optimization and learning, consistency between the learned policy and the optimization objective is crucial, as long-term performance depends on the LVQ model’s approximation of optimal user–cell associations. Associations are computed via Lagrangian relaxation, with coupling constraints managed by dual variables updated through subgradient iterations. For diminishing step sizes satisfying $\sum_{t=1}^{\infty}\alpha_t = \infty$ and $\sum_{t=1}^{\infty}\alpha_t^2 < \infty$, the dual variables converge to the optimal solution. Despite the discrete nature of decisions, this relaxation provides an efficient and widely used approximation. Simulations show rapid convergence, yielding stable supervisory signals for LVQ training. Denoting optimal and learned decisions at interval $t$ by $a_t^{opt}$ and $a_t^{ml}$, the approximation error
$E = \frac{1}{T}\sum_{t=1}^{T}\mathbb{1}(a_t^{ml} \neq a_t^{opt})$
quickly stabilizes at a low value, while cumulative regret
$R_T = \sum_{t=1}^{T}(J(a_t^{ml}) - J(a_t^{opt}))$
remains bounded, indicating that the learning module retains most optimization gains. The hybrid framework thus achieves near-optimal decision quality with scalable, real-time applicability in dense small-cell networks.

\vspace{-0.1in}
\section{Conclusion}
Dense small-cell deployments are expected to play a key role in future 6G networks but significantly increase the complexity of user association under dynamic traffic and mobility. This paper presented a KDN-based framework that combines Lagrangian optimization with a lightweight LVQ learning module for scalable real-time small-cell selection. By incorporating queue state and service-degradation indicators, the approach enables load-aware and QoS-driven association decisions. Simulation results from an NS-3 dense-network scenario show consistent gains over signal-based and heuristic baselines. The proposed method reduces average latency by up to 30--45\% under heavy traffic while decreasing packet loss by more than 35\%. More than 95\% of users meet the 50~ms latency target and over 90\% remain below the 2\% packet-loss threshold, demonstrating the effectiveness of combining optimization with lightweight learning for QoS-aware association in dense 6G networks.
\vspace{-0.1in}

\section{Future Work}
Despite the achieved performance gains, several limitations remain. The evaluation is conducted in a controlled NS-3 simulation environment with synthetic traffic. Although this ensures reproducibility, real deployments exhibit irregular traffic bursts, hardware heterogeneity, and complex mobility patterns. Future work will therefore include validation using real network traces and large-scale experimental platforms.

The current energy-awareness model reflects relative cell-level energy costs but does not explicitly model hardware characteristics, sleep modes, or detailed power-control mechanisms. Incorporating more realistic energy models will improve practical relevance. Moreover, the architecture assumes a single administrative domain with centralized knowledge-plane coordination. Since future 6G systems are expected to span multi-operator and multi-domain environments, federated and distributed KDN architectures will be investigated to enable collaborative knowledge exchange while limiting inter-domain signaling overhead.

Planned extensions include adaptive model updating to refine LVQ prototypes under evolving traffic, including online learning and concept-drift detection. Integration with network slicing and service-aware orchestration will be explored to jointly support heterogeneous QoS demands of URLLC, eMBB, and massive IoT services. Finally, tighter cross-layer integration with radio resource management and semantic-aware networking paradigms will be studied to further enhance scalability and intelligence in future 6G systems.

\bibliographystyle{IEEEtran}
\bibliography{ref}

\end{document}